\documentclass[aps,pra,float,onecolumn,superscriptaddress]{revtex4-2}
\usepackage{graphicx}
\usepackage{braket}
\usepackage{amsmath}
\usepackage{dsfont}
\usepackage[colorlinks, citecolor=blue, urlcolor=red, linkcolor=blue]{hyperref}
\usepackage{amsfonts}
\usepackage{float}
\usepackage[font=small,format=plain,labelsep=period,labelfont=normal,textfont=normal,singlelinecheck=false,justification=raggedright]{caption}
\usepackage{subcaption}
\usepackage{xcolor}
\usepackage{comment}
\usepackage{textcomp}
\usepackage{amsthm}
\usepackage{amssymb}
\usepackage{booktabs}
\usepackage[left=1.3cm,right=1cm,top=2cm,bottom=2.5cm]{geometry}

\begin{document}

\title{Quantum walks in polycyclic aromatic hydrocarbons}

\author{Prateek Chawla}
\email{prateekc@imsc.res.in}
\affiliation{The Institute of Mathematical Sciences, C. I. T. Campus, Taramani, Chennai 600113, India}
\affiliation{Homi Bhabha National Institute, Training School Complex, Anushakti Nagar, Mumbai 400094, India}
\author{C. M. Chandrashekar}
\email{chandru@imsc.res.in}
\affiliation{The Institute of Mathematical Sciences, C. I. T. Campus, Taramani, Chennai 600113, India}
\affiliation{Homi Bhabha National Institute, Training School Complex, Anushakti Nagar, Mumbai 400094, India}
\affiliation{Department of Instrumentation \& Applied Physics, Indian Institute of Science, Bengaluru 560012, India}

\begin{abstract}
	Aromaticity is a well-known phenomenon in both physics and chemistry, and is responsible for many unique chemical and physical properties of aromatic molecules. The primary feature contributing to the stability of polycyclic aromatic hydrocarbons is the delocalised $\pi$-electron clouds in the $2p_z$ orbitals of each of the $N$ carbon atoms. While it is known that electrons delocalize among the hybridized $sp^2$ orbitals, this paper proposes quantum walk as the mechanism by which the delocalization occurs, and also obtains how the functional chemical structures of these molecules arise naturally out of such a construction. We present results of computations performed for some benzoid polycyclic aromatic hydrocarbons in this regard, and show that the quantum walk-based approach does correctly predict the reactive sites and stability order of the molecules considered.
\end{abstract}
	

\maketitle

\section{Introduction}
\label{sec:intro}
	
The structure and properties of Benzene and other arenes has been a subject of special interest in the field of quantum chemistry for a long time. It is now known that each of the six carbon atoms in Benzene have their orbitals hybridized into the $sp^2$ state, and the $2p_z$ orbitals host a delocalized $\pi$-electron cloud of the molecule. This type of cyclic conjugated system has been a subject of significant interest in the chemical sciences, and the description of such systems is studied under the concept(s) of (anti)aromaticity \cite{PG01,R03, KCY05,MMM08,GH12}. These molecules are characterized by their electron-rich clouds, which stabilize the structure through cyclic resonant structures. While it is known that the delocalization does occur in the electron cloud, the underlying physical process by which the process takes place is an open question as of yet. 

In order to understand the chemical characteristics of the cyclic conjugated system, bond order of the bonds in the system is studied. Bond order is a widely used concept in chemical sciences, and works as a tool to help predict chemical behaviour. Certain reactions can take place only for bonds of certain order (e.g. electrophilic and nucleophilic substitution reactions can only occur for bond order $\geq 1$). It can also help in explaining the reactive tendencies of some molecules by consideration of the sum of bond order \cite{DG89} of the atoms preferred by the molecule. The strength of a chemical bond depends on two things - the degree of overlap between the interacting orbitals, and the difference in energies of the atomic orbitals involved in bonding, which is reflected by bond polarity. The bond polarity is used to qualify the ionic nature of the bond, while the degree of overlap quantifies its covalent nature \cite{P60,M50}. 

A comprehensive method to calculate bond order from first principles was only available in 2017 \cite{M17}. As we investigate polycyclic aromatic rings in this work, we can use a slightly simpler method that relies on the system's geometry. It requires a basis relative to which bond order can be calculated. We make the traditional choice of $C-C$ single bond in Ethane ($H_3C-CH_3$) to have bond order 1, and the $C=C$ double bond of Ethene ($H_2C=CH_2$) to have a bond order of 2. This gives us a quantity known as Relational Bond Strength Order (RBSO) \cite{FKC12}, and provides a robust description of bond strength as long as the Born-Oppenheimer approximation is valid. This provides a {useful tool to study the bonds in an aromatic molecule}, as a bond with bond order closer to $1$ would have characteristics that are closer to a single bond, but in a bond with order closer to $2$, the double-bond characteristics will dominate.

{With advancements in the field of quantum information and computation, a quantum walk has emerged as one of the most efficient ways to model the controlled dynamics of quantum states and quantum particles \cite{GVR58,RPF86,ADZ93,DAM96, FG98}. The quantum walk can be regarded as a quantum counterpart of random walks which have served as an efficient way to model dynamics defined by classical physics.  Therefore, it is natural to explore the potential of quantum walk to model quantum dynamics in the range of physical systems where quantum physics plays a defining role.   It is described in two main formalisms -- the discrete-time quantum walk (DTQW) and the continuous-time quantum walk (CTQW). The dynamics of DTQW require a coin and a position Hilbert space in order to be defined, however, CTQW dynamics can be defined with only a position Hilbert space. A quantum walk spreads quadratically faster than a classical walk on its position space and its dynamics can be engineered using a set of evolution parameters, and therefore it has been used as a basis for design and implementation of quantum algorithms and quantum simulations \cite{YKE08,JK03,ESV12, V12, ANAV17,IKS05, SF92, MRL08, KRBD10, C11, C13, MMC17, CMC20, PMCM13}, to study problems  such as graph isomorphism \cite{DW08},  quantum percolation \cite{CB14,KKNJ12,CAC19}, and to develop schemes for implementation of universal quantum computation \cite{SCSC19, SCASC20}, among others. Quantum walks are thus very versatile tools indeed, and their practical significance has been demonstrated by way of implementation in many quantum systems, such as NMR\,\cite{RLBL05}, integrated photonics\,\cite{SCP10,BFL10,P10}, ion traps\,\cite{SRS09,ZKG10}, and cold atoms\,\cite{ KFCSWMW09}.}

{Modelling the dynamics of the FMO complex using quantum walks is one of the important works reported \cite{MRL08,GC20} in the direction of using quantum walks for modelling quantum dynamics in physical systems. However, not much progress has been reported in modelling of dynamics in chemical systems beyond FMO complex despite various algorithms having been proposed for using quantum walks for several computational tasks. Quantum simulation of dynamics in chemical complexes is one of the promising applications envisioned using quantum computers, and quantum algorithms using quantum walks could play an important role in that direction.}

In this work, {we will make use of  both the discrete-time and continuous-time quantum walk formalisms} in order to study the structure, stability, and the relative chemical reactivity of different sites when exposed to an electrophile for each of the molecules considered. We also qualitatively establish that the order of stability of the molecules arises naturally from a quantum walk-based framework. We thus propose the hypothesis that the electrons in the $2p^z$ orbitals delocalize in the $\pi$-electron cloud via a quantum walk.  

{The} paper is divided into six sections. In Sec.~\ref{sec:ctqw}, we discuss the discrete- and continuous-time quantum walks, and elucidate briefly the specific variants of the quantum walks used in this work. A quick recap of the concepts of relative bond strength order is given in Sec.~\ref{sec:rbso}, and the {methods we use to model the dynamics in aromatic hydrocarbons} are detailed in Sec.~\ref{sec:method}. We present our results in Sec.~\ref{sec:res}, and conclude in Sec.~\ref{sec:conc}.

\section{Quantum walk}
\label{sec:ctqw}

\subsection{Continuous-time quantum walk}

A quantum walker performing CTQW has its position space defined by a graph $\Gamma(V,E)$, where $V$ is the set of its vertices and $E$ the set of edges. Given the cardinality of $V = |V| = N$, $|E| \leq \frac{N(N-1)}{2}$. Let $A$ be the matrix defined as,

\begin{equation}
\label{eq:eq2.1}
		A_{ij} := \begin{cases}
		1 & \text{edge } (i,j) \in E \\
		0 & \text{otherwise}
	\end{cases}.
\end{equation}
$A$, defined as such on $\Gamma$, is known as the adjacency matrix of the graph, and offers a representation of $\Gamma$ itself. It is a real-valued matrix, and the vertices are labeled by the computational basis states $\{\ket{1}, \ket{2}, ..., \ket{N}\}$. The quantum state of the entire graph is defined at an arbitrary point of continuous time $t$ by the wavefunction $\ket{\psi(t)}$, which is defined as,

\begin{equation}
	\label{eq:eq2.2}
	\ket{\psi(t)} = \sum_{l=1}^{N} \alpha_l\ket{l} \;\;\;\; \alpha_l \in \mathbb{C}.
\end{equation}
{The CTQW is a quantum process, and thus obeys the Schr\"{o}dinger equation,
\begin{equation}
	\label{eq:eq2.3}
	i\hbar \frac{\partial}{\partial t} \ket{\psi(t)} = H_\Gamma \ket{\psi(t)},
\end{equation}
 while its classical counterparts obey the Markovian master equation. The Hamiltonian $H_\Gamma$ used in the expression of the Schr\"{o}dinger equation is given by,}
\begin{equation}
	\label{eq:eq2.4}
	\begin{aligned}
		H_\Gamma  &= \gamma L = \gamma(D-A), \\
		\implies H_{\Gamma_{ij}} &= \begin{cases}
			-\gamma & i\neq j,\;\; (i,j) \in E \\
			0 & i\neq j,\;\; (i,j) \notin E \\
			d_{ii}\gamma & i = j,
		\end{cases} 
	\end{aligned}.	
\end{equation}
Here $D$ is a diagonal matrix, and each diagonal element $d_{jj}$ corresponds to the degree of vertex $j$. $\gamma$ is a constant, $A$ is the adjacency matrix (defined as in Eq.\,(\ref{eq:eq2.1})), $L$ is known as the Laplacian of $\Gamma$, and $\gamma$ is the rate of transition for the graph. 

It may be observed from the expression in Eq.\,(\ref{eq:eq2.4}), that $H_\Gamma$ is independent of time, and thus the solution to Eq.\,(\ref{eq:eq2.3}) is given by the time evolution operator $U$, and may be expressed as,
\begin{equation}
	\label{eq:eq2.5}
	\begin{aligned}
	U &= e^{-iH_\Gamma t}, \\
	\ket{\psi(t)} &= U \ket{\psi(0)},
	\end{aligned}
\end{equation}
\noindent
where the expression of $\psi$ is in the units of $\hbar$. Since the adjacency matrix $A$ is always real and symmetric, due to it representing an undirected graph, so is the Hamiltonian - which ensures that $U$ is necessarily a unitary operation.

The coefficients $\gamma$ in the Hamiltonian in Eq.\,\eqref{eq:eq2.4} are rates of transition between different nodes. Traditionally, in unweighted graphs, the rates are all normalized to $1$. In the case of arenes, however, {we use the values of RBSO as mentioned in the beginning of this paper in Sec.~\ref{sec:intro}, and formally defined in Sec.~\ref{sec:rbso}} as the values of the weights of the graph. This corresponds to how easily the electrons are able to move across the bonds, and physically represents a measure of the bond length, as bonds of higher order are shorter.

\subsection{Discrete-time quantum walk}

The quantum evolution of a walker executing a DTQW on a one-dimensional lattice may be described on a Hilbert space $\mathcal{H} = \mathcal{H}_C \otimes \mathcal{H}_P$, where $\mathcal{H}_C$ and $\mathcal{H}_P$ are coin and position Hilbert spaces, respectively. The coin space in this case is a 2-dimensional space, and thus its basis set can be assumed to consist of two vectors $\ket{\uparrow}, \ket{\downarrow}$, such that they are mutually orthogonal. The coin space represents a Hilbert space which is internal to each walker. The position space is an infinite-dimensional Hilbert space with the basis chosen to be the set $ \set{ \ket{x} | x \in \mathbb{Z} }$. Each vector $\ket{n}$ represents the site $x=n$ in the position space.

The initial state of the walker is therefore represented as a tensor product of its states in the two spaces, and may be written in the form shown in Eq.\,(\ref{eq:eq2.6}).
\begin{equation}
	\label{eq:eq2.6}
	\ket{\psi(0)}  = \left( \alpha \ket{\uparrow} + \beta \ket{\downarrow}\right) \otimes \ket{x=0} \; , \hspace*{5pt} \text{where }  |\alpha|^2 + |\beta|^2 = 1.
\end{equation}
Here, $\alpha,\beta \in \mathbb{C}$ are amplitudes of the walker's internal coin states. The evolution is defined as a unitary operation in the coin space, followed by a coin-dependent shift operator in the position space. Both the operators are unitary operations, and a typical DTQW evolution with a single-parameter coin is described as,
{\begin{equation}
	\label{eq:eq2.7}
	\begin{aligned}
		\ket{\psi(t)} &= \left( S_x \left(C_\theta \otimes \mathds{1_{pos}}\right) \right)^t \ket{\psi(0)}, \\
		\text{where,} &{} \\
		C_\theta &= \begin{bmatrix}
			~~~\cos(\theta) & -i\sin(\theta) \\
			-i\sin(\theta) & ~~~\cos(\theta)
		\end{bmatrix}, \text{ and} \\
		S_x &= \sum_{x \in \mathbb{Z}} \bigg[ \ket{\uparrow}\bra{\uparrow} \otimes \ket{x\pm a} \bra{x} + \ket{\downarrow}\bra{\downarrow} \otimes \ket{x\pm b}\bra{x} \bigg],
	\end{aligned}	
\end{equation}}
\noindent
{Where $(a,b) \in \mathbb{R}$ represent the amount of shifting in the position space experienced by the component in the eigenspaces corresponding to $\ket{\uparrow}$ and $\ket{\downarrow}$ respectively.}
 
As can be seen from Eq.\,(\ref{eq:eq2.7}) the shift operation effects the traversal of the components of the probability amplitude in different directions. In order to obtain the order of reactivity of sites, the algorithm in \cite{CMC20} requires a variant of DTQW known as a directed DTQW (D-DTQW) \cite{HM09}. {In case of the D-DTQW, the shift operator only allows a single component of the probability amplitude to traverse the graph. Thus depending on which component is allowed to traverse the graph, the D-DTQW shift operation may be defined in one of the two ways,}
\begin{equation}
	\label{eq:eq2.8}
	S_\pm = \begin{cases}
		\sum_x \ket{\uparrow}\bra{\uparrow} \otimes \ket{x\pm 1}\bra{x} + \ket{\downarrow}\bra{\downarrow} \otimes \ket{x}\bra{x} \\
		\sum_x \ket{\uparrow}\bra{\uparrow} \otimes \ket{x}\bra{x} + \ket{\downarrow}\bra{\downarrow} \otimes \ket{x\pm 1}\bra{x}.
	\end{cases}
\end{equation}

{The coin operation for the node-ranking algorithm has the same form as $C_\theta$, however, it uses a position dependent form which may be expressed as,
\begin{equation*}
	\hat{C}(\theta)_x = \sum_x C(\theta) \otimes \ket{x}\bra{x}.
\end{equation*}
This operation is also a special unitary matrix, similar to the $C_\theta$ defined for the DTQW in Eq.\,(\ref{eq:eq2.7}). In case of the procedure to arrange the nodes in order of reactivities, {we use a specialized coin operation of the form defined in Ref.\,\cite{CMC20},}}
\begin{equation}
	\label{eq:eq2.9}
	C = \sum_x \begin{bmatrix}
		\sqrt{\frac{1}{\alpha_x+1}} & \sqrt{\frac{\alpha_x}{\alpha_x+1}} \\
		\sqrt{\frac{\alpha_x}{\alpha_x+1}} & -\sqrt{\frac{1}{\alpha_x+1}}
	\end{bmatrix} \otimes \ket{x}\bra{x},
\end{equation}
\noindent
were $\alpha_x$ is the proportion of the incoming weight with respect to the total incoming and outgoing weights at the node represented by $\ket{x}$. In this case, the incoming and outgoing weights are equal as the graph is undirected, hence the expression for $\alpha_x$ may be simplified to $\alpha_x = \frac{d_x}{2}$, where $d_x$ represents the degree of node $\ket{x}$ .

{The shift operator used to implement the algorithm is defined as 
\begin{equation}
	\label{eq:eq2.10}
	S_{NR} = \sum_{x \in \mathbb{Z}} \bigg[ \ket{\uparrow}\bra{\uparrow}\otimes \ket{x}\bra{x} + \sum_k \big[ U_{kx} \ket{\downarrow}\bra{\downarrow} \otimes \ket{k}\bra{x} \big] \bigg],
\end{equation}
\noindent
where $k\in \mathbb{Z}$, and $U_{kx}$ is a unitary matrix that restricts the walker to Markovian jumps between specific nodes in the position space. Since the coin operator rotates between the `stored ($\psi_c^\uparrow$)' and `travelling ($\psi_c^\downarrow$)' components of the probability distribution $\psi_c = \begin{bmatrix} \psi_c^\uparrow \\ \psi_c^\downarrow \end{bmatrix} $ , it results in different amounts of storage at different nodes, depending on the amount of information that passes through them, generating a ranking of the nodes. The position space in this case is the graph $\Gamma(V,E)$ with the set of nodes $V$ and edges $E$. The adjacency matrix $A$ of this graph is defined as $a_{ij} = RBSO(i,j)$, where $RBSO(i,j)$ is the relative bond strength order of the bond between the $i^{th}$ and $j^{th}$ carbon atoms in the molecule under consideration, as defined in Sec.~\ref{sec:rbso}.}

\section{Relative Bond Strength Order}
\label{sec:rbso}

Many computational approaches to calculate bond order exist, however, most have severe fundamental limitations. Approaches requiring categorization of electrons to be spin-up or spin-down fail to achieve universality as they cannot describe noncollinear spin magnetism. Some methods consider bond order to be an explicit functional of the total electron density and spin magnetization density functions. However, in the limit of a complete basis set, the density matrix is an overcomplete representation of the distribution of electron density. Thus, a functional of the density matrix may not necessarily be a functional of the electron density. This causes the bond order results to be inconsistent across different quantum chemistry methods, even if they yield the same electron density, spin magnetization density, and energy \cite{C63,M03}. {This dependence renders this particular kind of formulation unphysical.} 

Other approaches that can be used in its stead include Wheatley-Gopal and Laplacian correlations between overlap and bond order \cite{WG12, LC13}, Mayer Bond Index \cite{MS04} and First Order Delocalization Index \cite{MSSD07} applied to density-based charge partitions, Natural bond orbital \cite{DS12}, Adaptive natural density partitioning \cite{ZB08, GDSB13}, among others. 

In this work, we shall consider electron delocalization in benzoid polycyclic aromatic systems, specifically the $C-C$ conjugated double bonds in Benzene, Naphthalene, Anthracene and Phenanthrene. Since all the bonds under consideration are $C-C$ bonds, we can use a relative bond strength order \cite{KKC14} (hence called bond order) here. We determine the bond order by using experimentally determined vibrational frequencies of the bonds \cite{S72,BSV72,BCWCGK79,SS07} to calculate the force constant matrix. We then find the local modes of vibration that correspond to specific bonds, and thus obtain local stretching modes \cite{ZKKC12,KKC13}, which can be used to design a sensitive measure of bond strength.

Briefly, the method proceeds as follows. We calculate the force constant matrix corresponding to the complete set of vibrational frequencies of the molecule under consideration by Wilson's GF method \cite{W41}. The internal degrees of freedom ($\Set{q_1,...,q_{3N-6}}$) describing the potential energy surface (PES) in an optimal manner are often nonlinear, so it is assumed that the displacements with respect to the internal coordinates are small. This enables linearization of the internal coordinates as the set $\Set{Q_i}$, where $i=1,..,3N-6$. A PES $V$ can be expanded in a Taylor series around its minima in terms of $\Set{Q_i}$, and the force derivative matrix $F$ is then given by the Hessian of $V$. The first term in the Taylor expansion is adjusted with the zero point energy and the second term vanishes due to the evaluation at minima. 

{
Thus, we obtain the relation},
\begin{equation}
	\label{eq:eq3.1}
	V \approx \frac{1}{2} \sum_{i,j=1}^{3N-6} F_{ij}Q_iQ_j,
\end{equation}
which can then be compared with the classical vibrational kinetic energy of the form,
\begin{equation} 
	\label{eq:eq3.2}
	2T = \sum_{i,j=1}^{3N-6} g_{ij}(\mathbf{q})\dot{Q}_i\dot{Q}_j,
\end{equation}
\noindent
where $g_{ij}$ is an element of the metric tensor of the internal curvilinear coordinates, and $\dot{A} \equiv \frac{\partial A}{\partial t}$. Evaluating the metric tensor $\mathbf{g}$ in the minimum $\mathbf{q}$ of the PES $V$ gives {the Wilson's G-matrix}, 
\begin{equation}
	\label{eq:eq3.3}
	\mathbf{G} = \mathbf{g}(\mathbf{q})^{-1}.
\end{equation}

This leads us to Wilson's equation, given by
\begin{equation}
	\label{eq:eq3.4}
	\mathbf{GFD} = \mathbf{D\Lambda},
\end{equation}
where $\mathbf{F}$ is the force constant matrix in terms of the internal coordinates ($\set{q_i}_{i=1}^{3N-6}$), $\mathbf{D}$ is the matrix of vibrational eigenvectors $\mathbf{d_\mu}$, each of which forms one of its columns, $\mathbf{G}$ is the Wilson matrix from Eq.~\eqref{eq:eq3.3}, and $\mathbf{\Lambda}$ is a diagonal matrix consisting of the eigenfrequencies (normal modes) corresponding to $\mathbf{d_\mu}$. 

Diagonalizing $\mathbf{F}$ with $\mathbf{D}$, we obtain the force constants corresponding to each local mode as,

\begin{equation}
	\label{eq:eq3.5}
	k_\mu = \mathbf{d_\mu \mathbf{K}^{-1} d_\mu^\dagger}, \hspace*{5pt} \text{where} \hspace*{5pt} \mathbf{K}=\mathbf{D^\dagger FD}.
\end{equation}

In this work, we use the vibrational frequencies as listed in \cite{S72} to calculate our force constants. According to the extended version of Badger rule \cite{KLC10}, the bond order can be calculated from a power relationship between it and local mode corresponding to each bond. We use the $CC$ bonds in Ethane and Ethene as having bond orders $1$ and $2$, respectively, which fixes the relationship between bond order and the local vibrational mode force constants as, \cite{KKC14}
\begin{equation}
	\label{eq:eq3.6}
	BO = 0.29909 (k_\mu)^{0.86585}.
\end{equation} 
	
It may be noted that the bond order of a bond that does not exist in the molecule is by definition considered to be zero. In Table~\ref{tab:tab3.1}, we show the bond order values corresponding to the different $C-C$ bonds as per Fig.~\ref{fig:fig3.1}.

\begin{table}[!h]
	\begin{minipage}[c]{0.75\textwidth}
		\centering
		\begin{tabular}{cc}
			\includegraphics[width=0.18\linewidth]{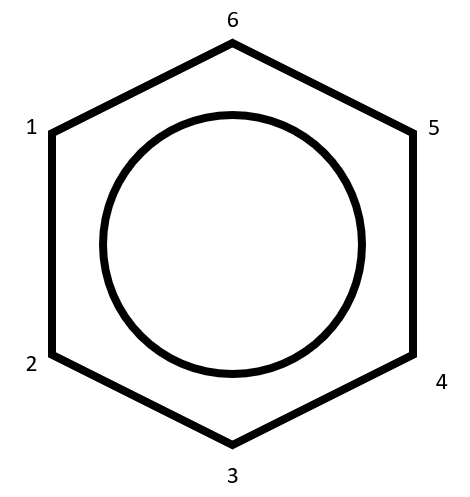} &
			\includegraphics[width=0.32\linewidth]{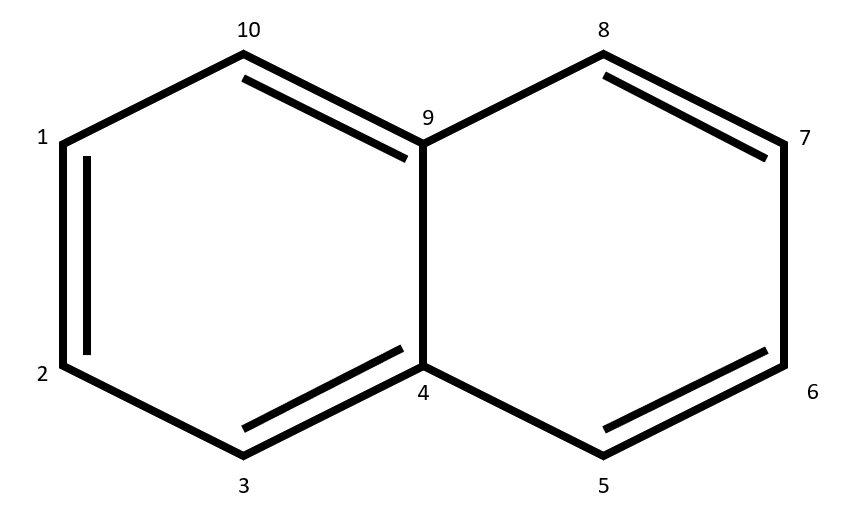} \\
			(a) & (b) \\
			\includegraphics[width=0.41\linewidth]{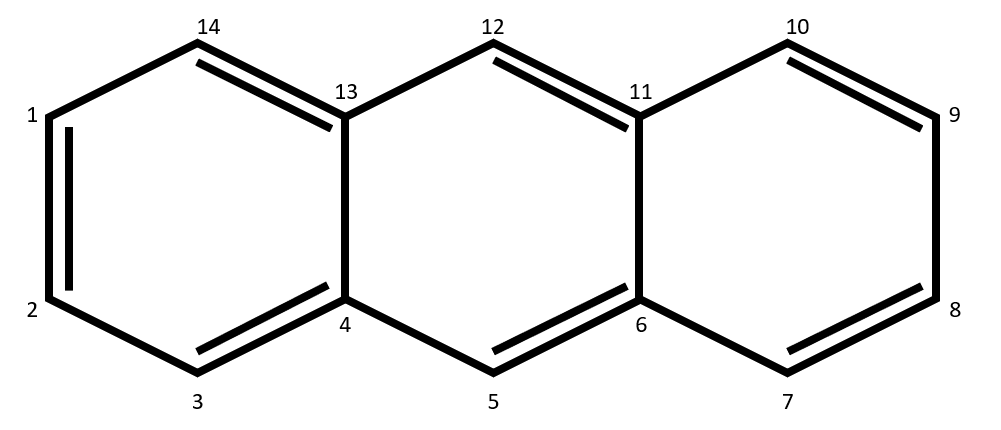} &
			\includegraphics[width=0.41\linewidth]{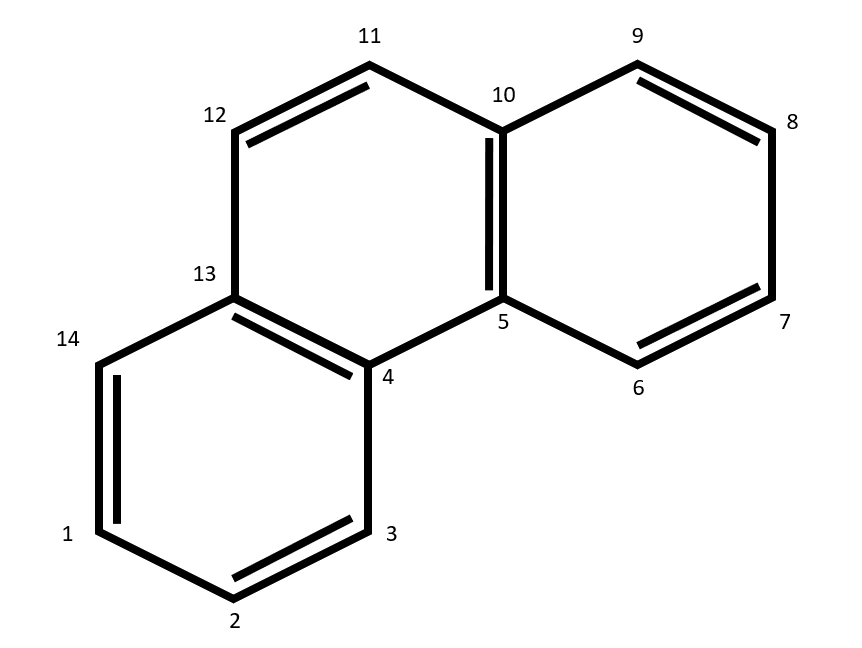} \\
			(c) & (d) \\
		\end{tabular}
	\captionof{figure}{
		A description of the benzoid polycyclic aromatic molecules considered in this study. The labels refer to the respective carbon atoms. The figures show the following molecules - (a) Benzene, (b) Naphthalene, (c) Anthracene, and (d) Phenanthrene.
		\label{fig:fig3.1}
	}
	\end{minipage}
\end{table}

\begin{table}[!h]
\centering
\renewcommand{\arraystretch}{1.2}
\squeezetable
\begin{tabular}{|cc|cc|} 
	\toprule[1pt]
	\multicolumn{2}{|c|}{\textbf{Benzene}} & \multicolumn{2}{c|}{\textbf{Naphthalene}}  \\ 
	\midrule[0.5pt]
	Carbon atoms & Bond order 
	& Carbon atoms & Bond order \\ 
	\midrule[1pt]
	C1,C2 & 1.468 & C1,C2 & 1.339  \\
	C6,C1 & 1.468 & C1,C10 & 1.603 \\
	C2,C3 & 1.468 & C2,C3 & 1.603  \\
	C3,C4 & 1.468 & C3,C4 & 1.335  \\
	C4,C5 & 1.468 & C4,C5 & 1.335  \\
	C5,C6 & 1.468 & C4,C9 & 1.288  \\
	    &       & C5,C6 & 1.603  \\
	    &   	& C6,C7 & 1.339  \\
	    &  		& C7,C8 & 1.603  \\
     	&  		& C8,C9 & 1.335  \\
	    &  		& C9,C10 & 1.335 \\
	\toprule[1pt]
	\multicolumn{2}{|c|}{\textbf{Anthracene}} & \multicolumn{2}{|c|}{\textbf{Phenanthrene}} \\
	\midrule[0.5pt]
	Carbon atoms & Bond order 
	& Carbon atoms & Bond order \\
	\midrule[1pt]
	C1,C2 & 1.295 & C1,C2 & 1.391 \\
	C1,C14 & 1.673 & C1,C14 & 1.571 \\
	C2,C3 & 1.673 & C2,C3 & 1.553 \\
	C3,C4 & 1.304 & C3,C4 & 1.348 \\
	C4,C5 & 1.452 & C4,C5 & 1.204 \\
	C4,C13 & 1.246 & C4,C13 & 1.315 \\
	C5,C6 & 1.452 & C5,C6 & 1.348 \\
	C6,C7 & 1.304 & C5,C10 & 1.315 \\
	C6,C11 & 1.246 & C6,C7 & 1.553 \\
	C7,C8 & 1.673 & C7,C8 & 1.391 \\
	C8,C9 & 1.295 & C8,C9 & 1.571 \\
	C9,C10 & 1.673 & C9,C10 & 1.367 \\
	C10,C11 & 1.304 & C10,C11  & 1.291 \\
	C11,C12  & 1.452 & C11,C12  & 1.762 \\
	C12,C13  & 1.452 & C12,C13  & 1.291 \\
    C13,C14  & 1.304 & C13,C14  & 1.367 \\
	\bottomrule[1pt]		
\end{tabular}
	\caption{Table showing the values of bond order obtained for various bonds in polycyclic aromatic molecules considered in this work. The atom numbers in the 'Carbon atoms' column correspond to the respective carbon atoms of the molecules, marked as shown in Fig.~\ref{fig:fig3.1}. \label{tab:tab3.1}}
\end{table}

\section{Methods}
\label{sec:method}

In this section, we consider the behavior of the $\pi$-electrons in the aromatic system. The behavior of the delocalised electrons in the conjugated system is modeled {in the form of quantum walk}. We will use both, the CTQW and DTQW formalisms to model the electrons' behavior. We will use the CTQW to model the delocalization process in order to obtain results relating to delocalization modes and stability of the molecule, {and the DTQW is employed in an algorithmic form defined in \cite{CMC20} to rank sites of the molecule in order of their reactivity towards an electrophile.}

In both the quantum walk formalisms, the molecules we have considered are modeled as a graph $\Gamma(V,E)$, where $V$ is the set of its nodes and $E$ the set of edges. Only the bonds and carbon atoms that form the conjugated system are included in the sets $E$ and $V$, respectively. The graph is represented in the form of its adjacency matrix, and the weight of edge is defined to be its bond order, as it represents the ease by which an electron may traverse the graph, thus qualitatively describing the ease by which an electronic wavefunction may delocalize over the bond. The graphs corresponding to each molecule are illustrated in Fig.~\ref{fig:fig3.1}. 

The $\pi$-electrons are modeled as independent, noninteracting quantum walkers that are free to delocalize over this graph. The interelectronic interactions between delocalized $\pi$ electrons in a polycyclic aromatic system are primarily nearest-neighbor pairwise interactions. {They are taken into consideration when the bond order is calculated, as the electrons find it easier to delocalize over a shorter bond, i.e., a bond with a higher bond order in general. Thus, the quantum walk dynamics of the electrons can be assumed }independent of each other, subject to the constraint that edge weights must take into account the effect of appropriate pairwise interactions along the network. 

In the CTQW study, we consider the state of the system after some time $t$, and look at the probabilities of finding some electron at a particular position and its variation with time. This gives us a general idea of the dynamics of the delocalizing electronic wavefunctions. We also look at how the maximum probability of any electron to exist at a particular position, hereafter called MAXP, varies over time. It is defined for each position basis vector as the maximum probability of a single electron to exist at that particular basis vector in the position space. A high mean value here indicates that one of the $\pi$-electrons is regularly found here, and therefore is evidence of localization, indicating that at least one of the $\pi$-bonds connecting the considered vertex has a higher bond order, i.e., it is not a dominant part of the delocalization mode of the molecule, and thus has a higher double-bond character. A lower value, on the other hand, corresponds to the fact that all the bonds at the considered vertex are involved in delocalization. This study helps to ascertain the bond delocalization mode of each molecule out of several possible choices, as illustrated in Fig.~\ref{fig:fig4.1}.


\begin{table}[!h]
	\begin{minipage}[c]{0.75\textwidth}
		\centering
		\begin{tabular}{ccc}
			\includegraphics[width=0.14\linewidth]{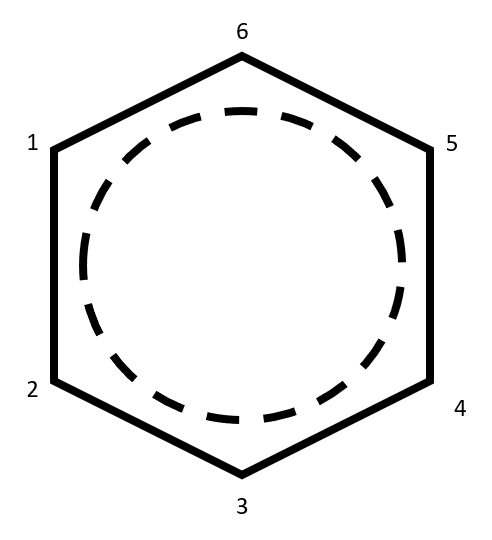} &
			\includegraphics[width=0.24\linewidth]{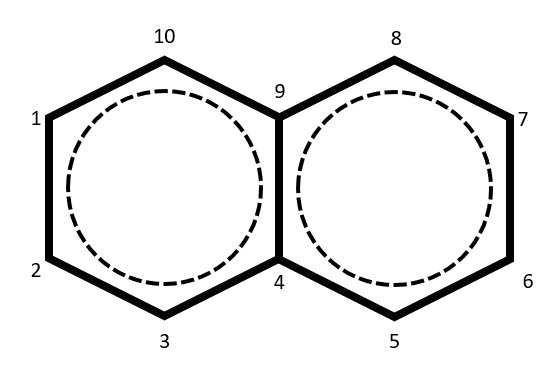} &
			\includegraphics[width=0.24\linewidth]{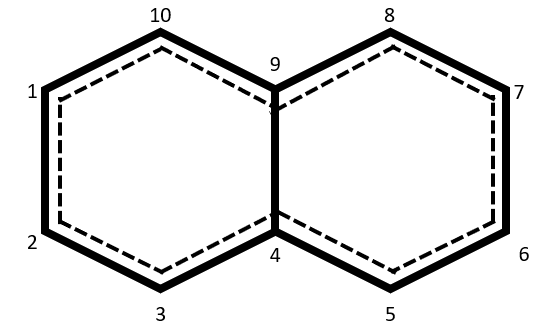} \\
			(a) & (b) & (c)\\
			
			\includegraphics[width=0.33\linewidth]{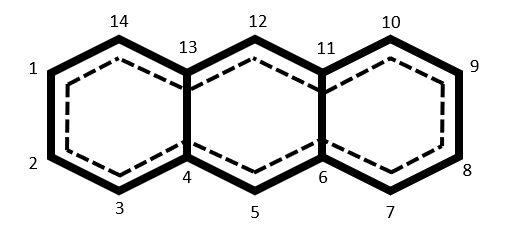} &
			\includegraphics[width=0.33\linewidth]{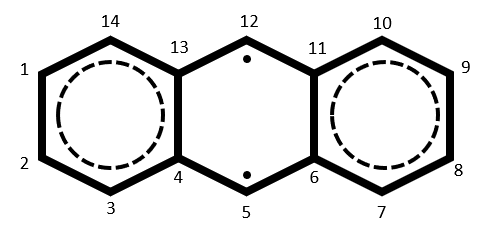} &
			\includegraphics[width=0.33\linewidth]{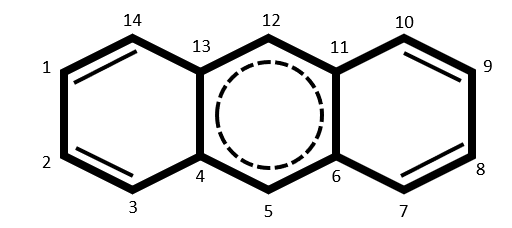} \\
			(d) & (e) & (f)\\
			
			\includegraphics[width=0.33\linewidth]{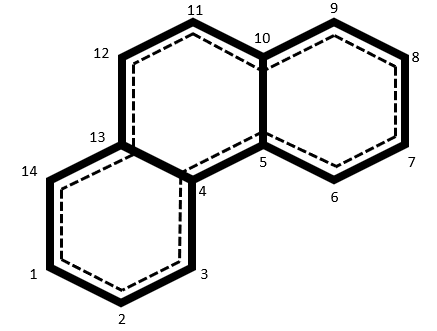} &
			\includegraphics[width=0.33\linewidth]{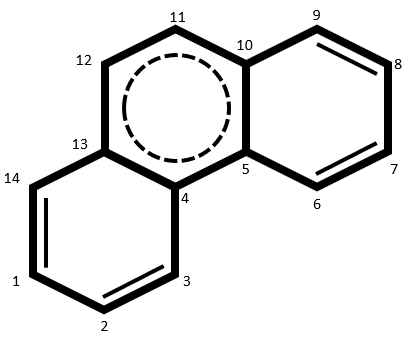} &
			\includegraphics[width=0.33\linewidth]{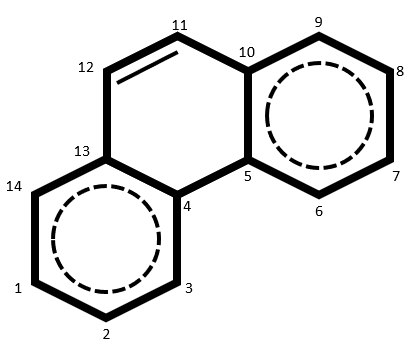} 
			\\ 
			(g) & (h) & (i)\\
		\end{tabular}
		\captionof{figure}{
			An illustration of the different possible delocalization modes for 
			each of the benzoid polycyclic aromatic hydrocarbons considered in 
			this work. Fig.~(a) illustrates the only possible delocalization 
			mode for Benzene, (b) and (c) depict the two possible ways by which 
			naphthalene can delocalize. The possibilities for Anthracene are 
			depicted in Figs.~(c), (d), and (e), and those for Phenanthrene are 
			represented in (g), (h), and (i). 
			\label{fig:fig4.1}
		}
	\end{minipage}
\end{table}

We also consider the truncated mean of this data as a variable called TRP, in which we discard the highest and lowest values of the probability, and calculate the mean of the remaining probabilities at a particular position over time. This measure discards the outlier(s) created by electrons that may localize, and provides a reasonable qualitative description of where the wavefunction is likely to exist. This provides a way to qualitatively characterize the reactivities of the various molecules, and also a way to arrange them in order of their reactivities. A higher value of TRP implies the electronic probability distribution often has a peak at certain places, which provides an estimate of the likelihood of an electron being present at a certain point in the chain -- thus qualitatively estimating the availability of that site to form chemical bonds.

The DTQW formalism characterizes the symmetries of the considered graph with respect to a diffusing quantum particle. The DTQW-based algorithm used \cite{CMC20} requires a coin Hilbert space {mapped to the connections in structure} in addition to the position Hilbert space for its implementation, and estimates the ability of each site to accumulate information as a quantum particle diffuses across it. This provides a qualitative estimate of site activities, and thus enables a qualitative overview of the behavior of the conjugated system in the presence of an electrophile.

\section{Results}
\label{sec:res}

\subsection{Evolution of probability with time}
In this subsection, we take a look at the evolution of the probability of each 
electron to exist at different points for each molecule as a function of time. 
This shows a oscillatory behavior symmetric about the initial state in the case 
of Benzene, one period of which is shown in Fig.~\ref{fig:fig5.1}.

\begin{widetext}
	\begin{table}[!h]
		\begin{minipage}[c]{0.95\textwidth}
			\centering
			\begin{tabular}{cc}
				\includegraphics[width=0.4\linewidth]{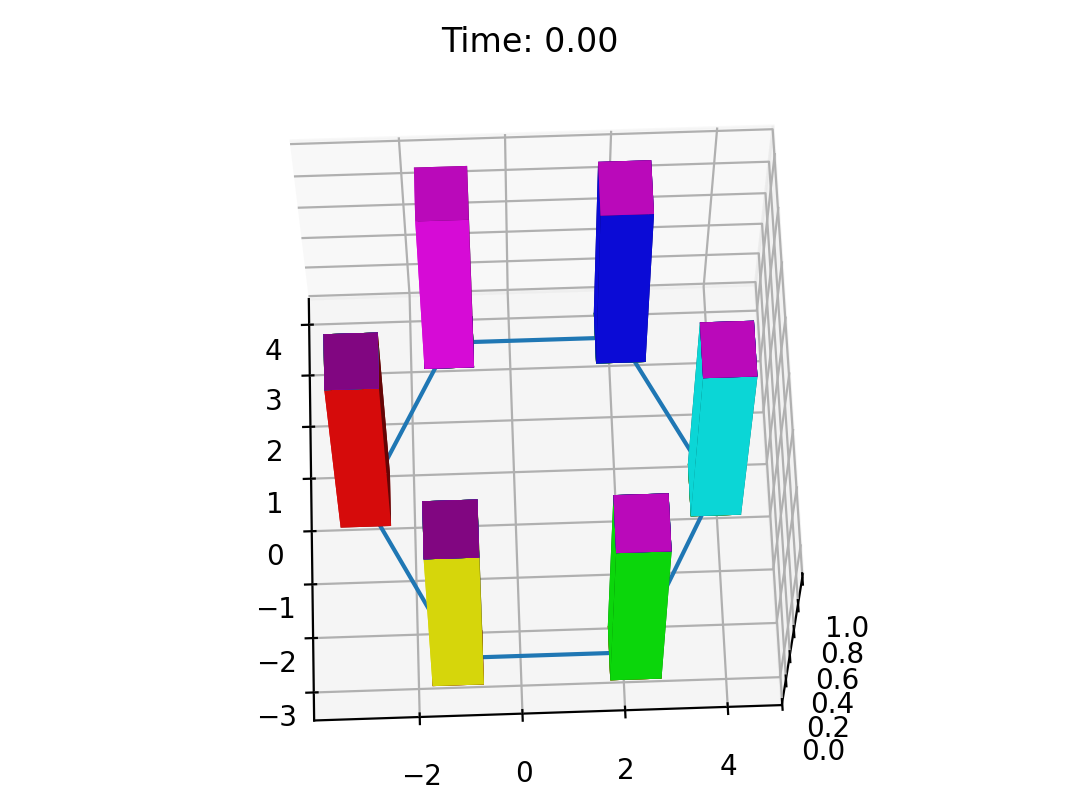} &
				\includegraphics[width=0.4\linewidth]{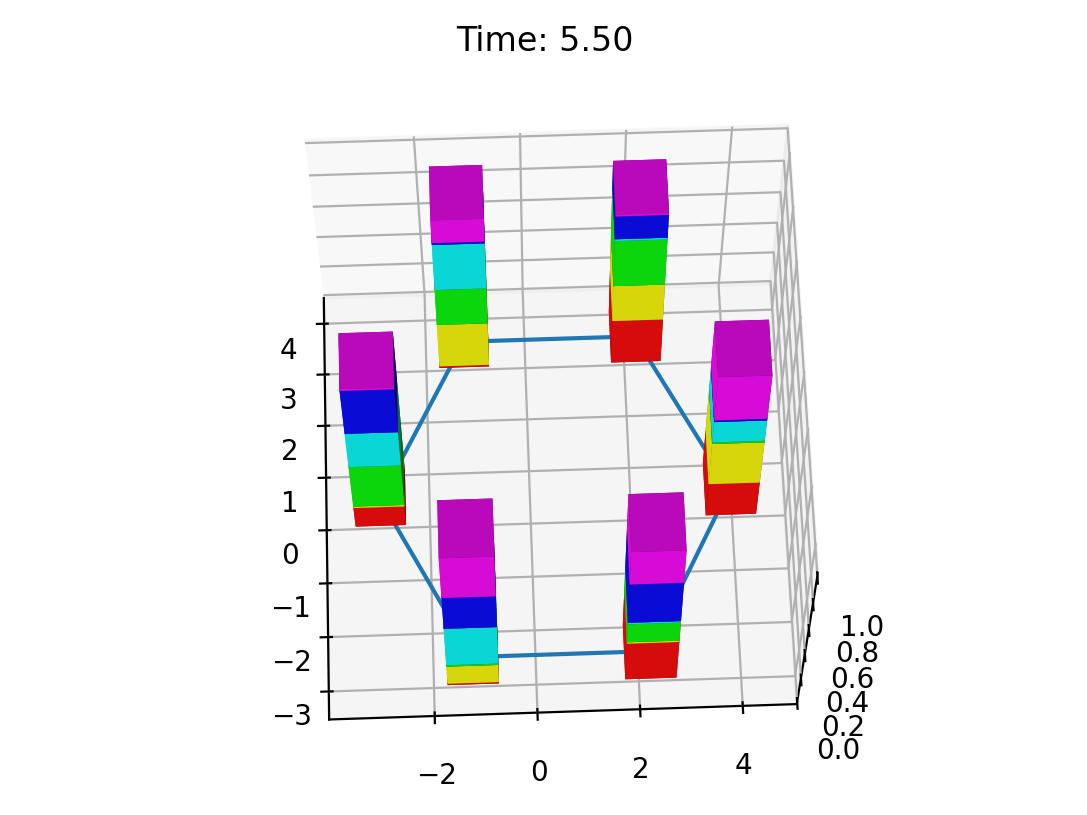} \\
				(a) & (b) \\
				\includegraphics[width=0.4\linewidth]{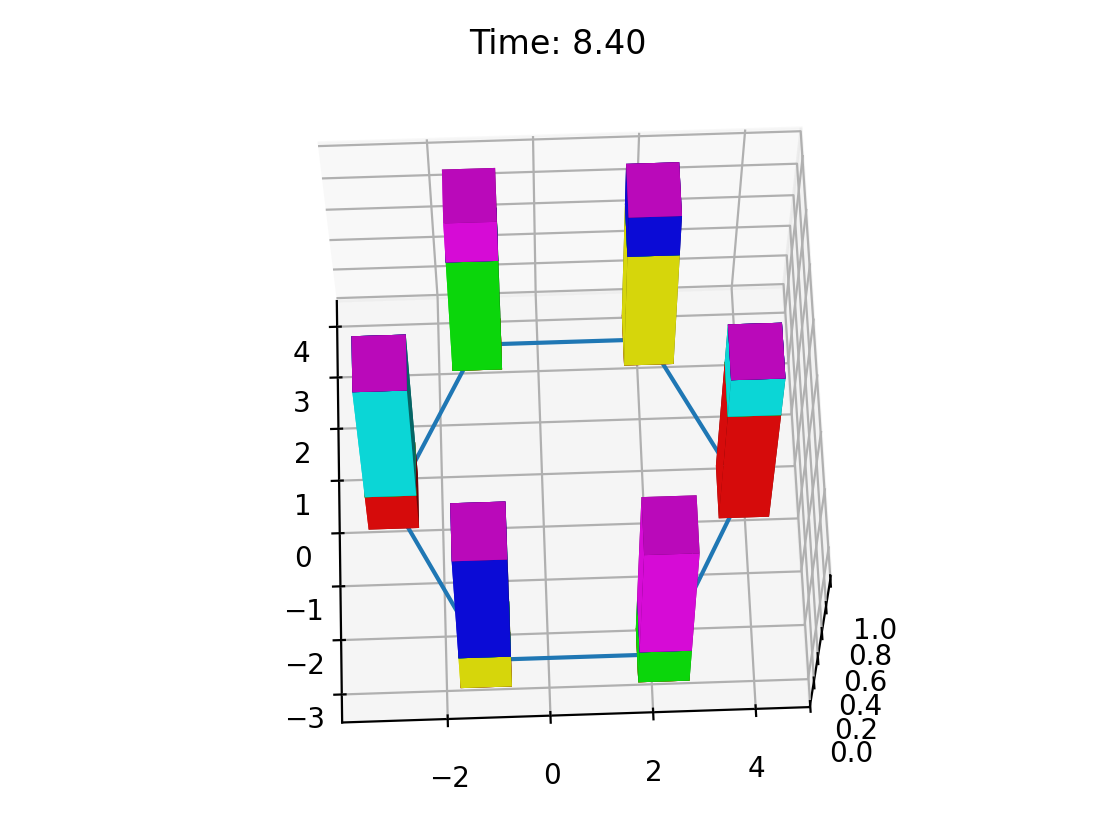} &
				\includegraphics[width=0.4\linewidth]{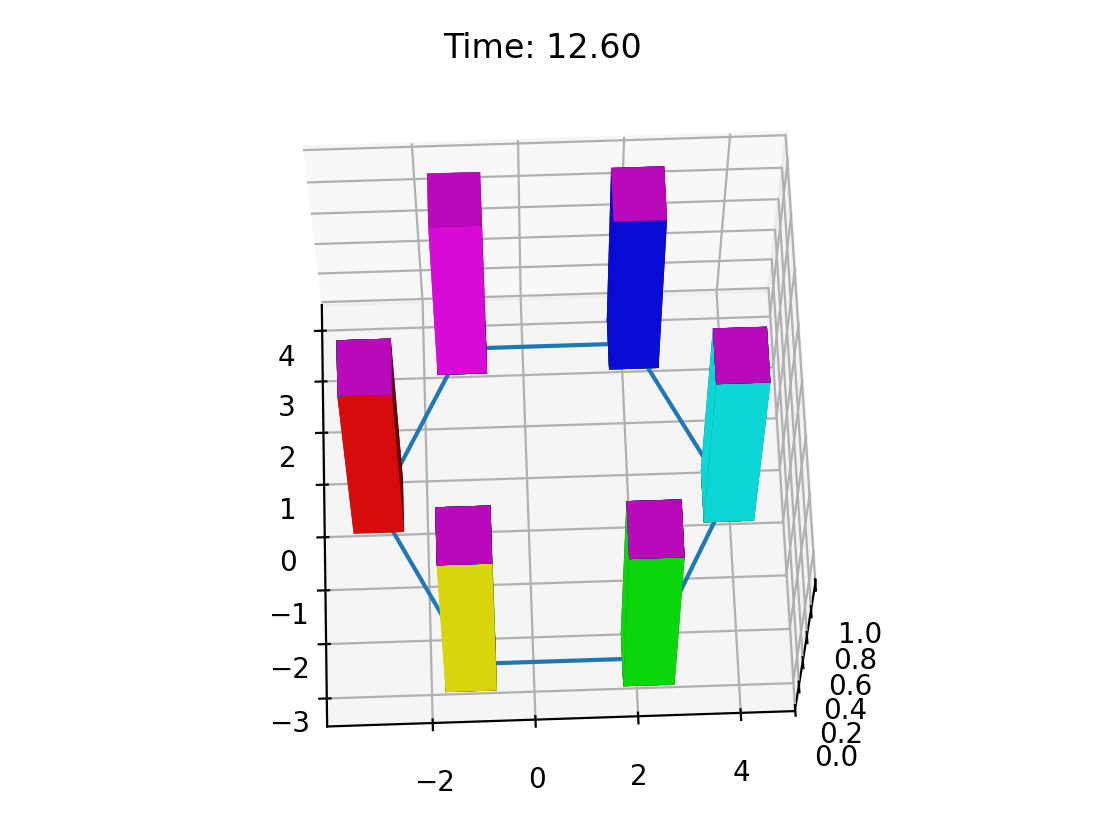} \\
				(c) & (d)\\
			\end{tabular}
			\captionof{figure}{
				The evolution of the probability distributions of the different walkers in the network representing Benzene, with the position basis superimposed on a plot of the Benzene ring for easier visualization. Each walker is represented by a different color. The evolution in case of Benzene exhibits an oscillation about the initial state.
				\label{fig:fig5.1}
			}
		\end{minipage}
	\end{table}
\end{widetext}

It is clear from Fig.~\ref{fig:fig5.1} that over time, the walkers diffuse on 
the network representing Benzene, as illustrated in Fig.~\ref{fig:fig3.1}(a) in 
an oscillatory manner, with a period of $12.60$. The total probability of 
finding an electron at a particular position is always unity, however, as can 
be seen from Fig.~\ref{fig:fig5.1}, it is always uncertain which electron is 
detected, in accordance with the concept of indistinguishability of electrons. 
The probabilities corresponding to each electronic wavefunction are also 
verified to sum to unity over all positions. This may be derived from the fact 
that the results for this calculation may be expressed as a bistochastic 
$N\times N$ matrix $B$. The $j^{th}$ row of $B$ represents the probability of 
the $j^{th}$ walker to be found at each of the $N$ sites.

\begin{widetext}
	\begin{table}[!h]
		\begin{minipage}[c]{0.95\textwidth}
			\centering
			\begin{tabular}{cc}
				\includegraphics[width=0.31\linewidth]{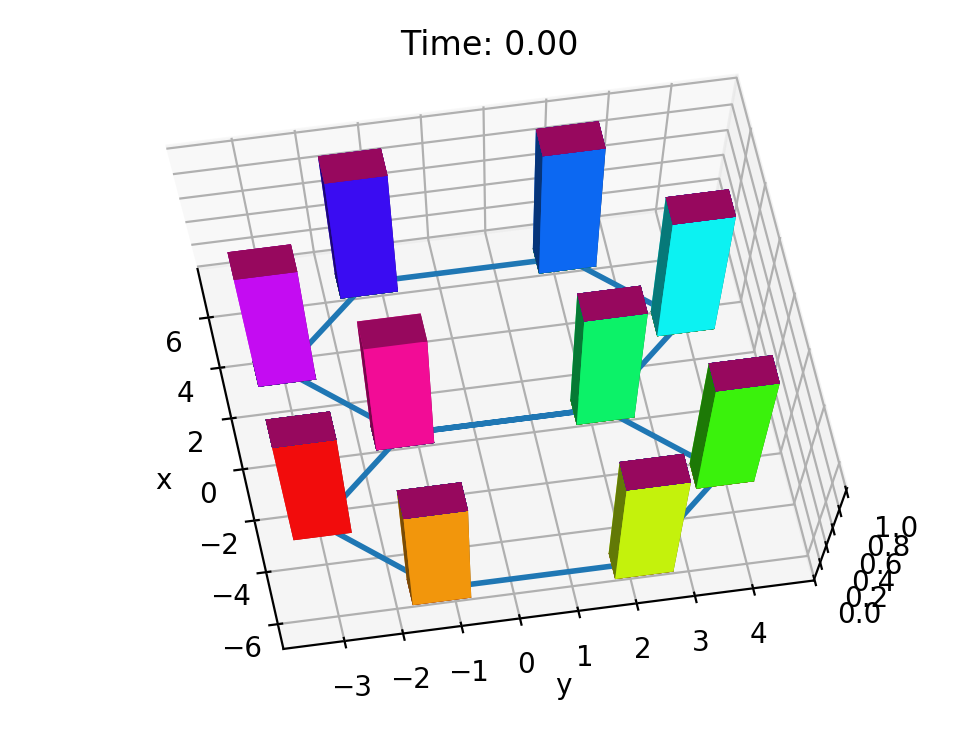} &
				\includegraphics[width=0.31\linewidth]{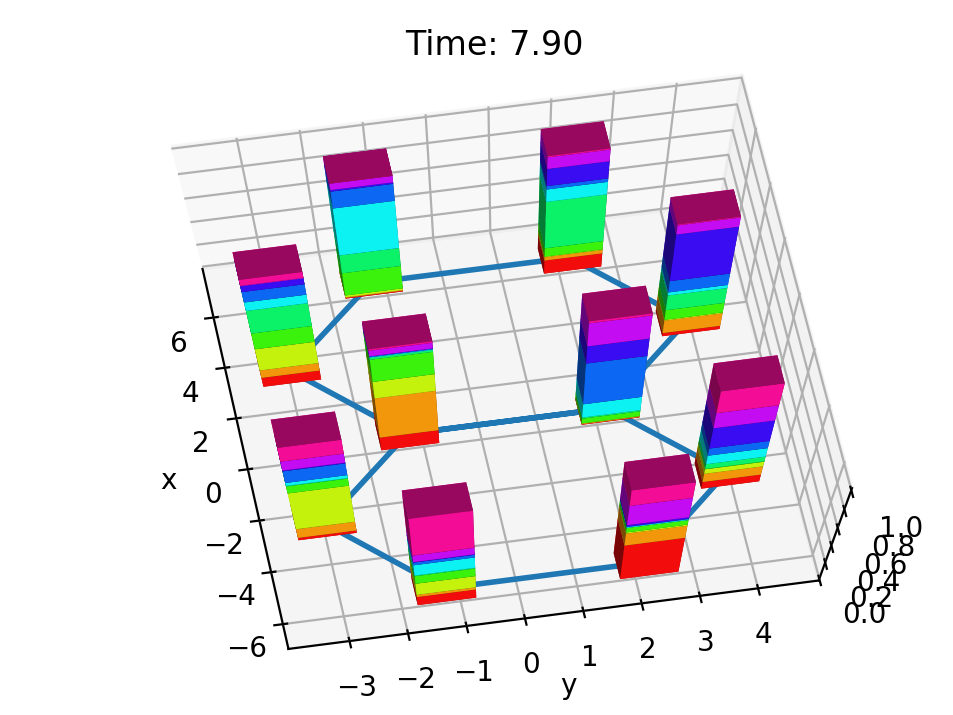} \\
				(a) & (b) \\
				\includegraphics[width=0.31\linewidth]{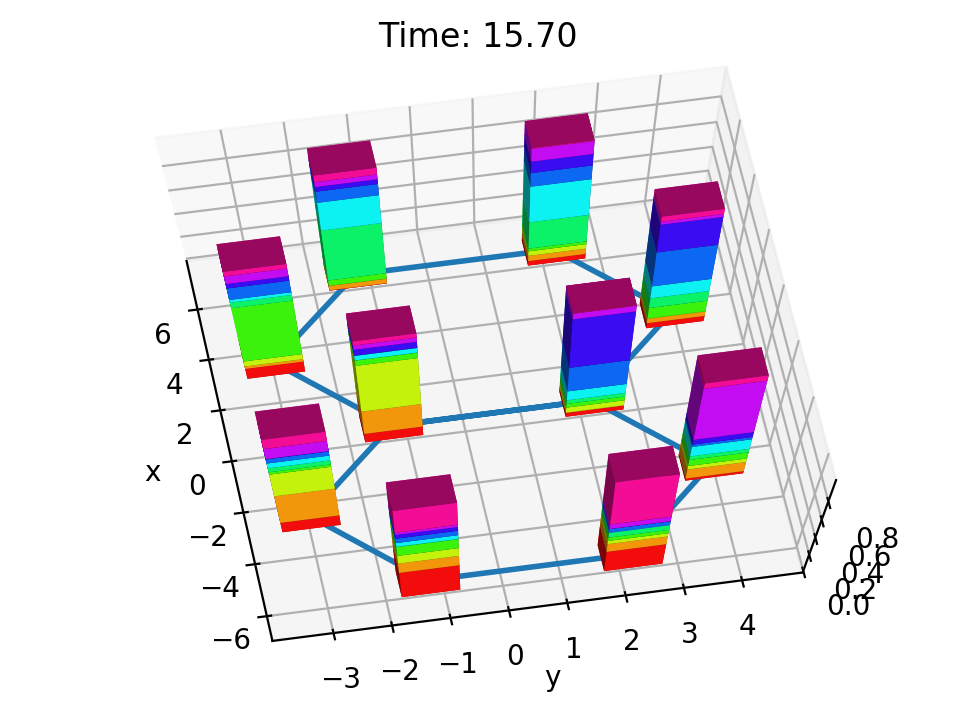} &
				\includegraphics[width=0.31\linewidth]{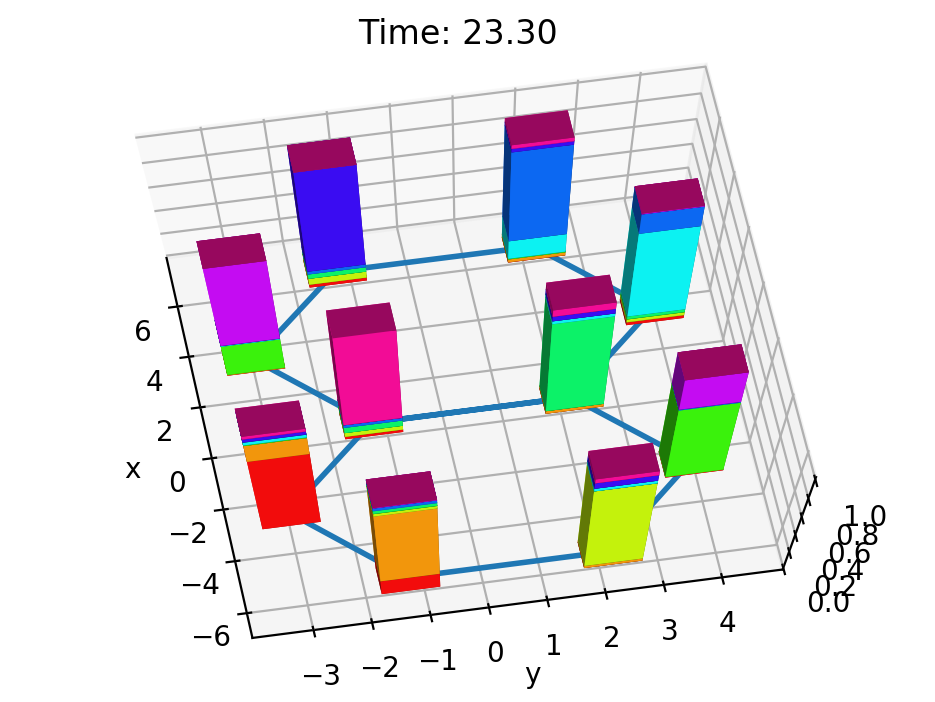} \\
				(c) & (d)\\
			\end{tabular}
			\captionof{figure}{
				The evolution of the probability distributions of the different walkers in the network representing Naphthalene, with the position basis superimposed on a representation of the network representing Naphthalene. Each walker is represented by a different color.
				\label{fig:fig5.2}
			}
		\end{minipage}
	\end{table}
\end{widetext}

\begin{widetext}
	\begin{table}[!h]
		\squeezetable
		\begin{minipage}[c]{0.95\textwidth}
			\centering
			\begin{tabular}{cc}
				\includegraphics[width=0.31\linewidth]{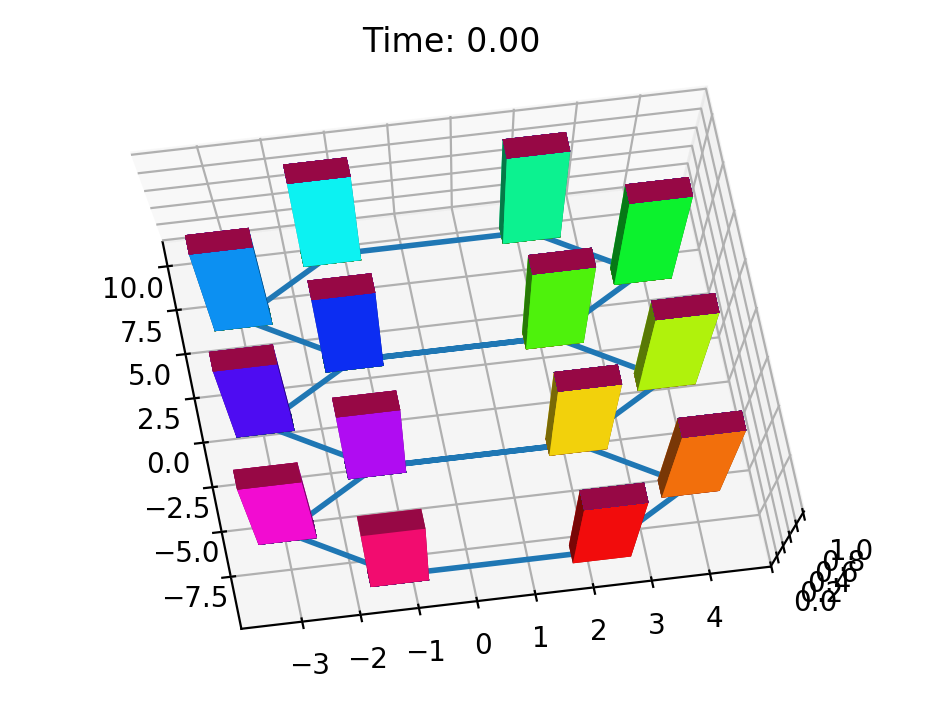} &
				\includegraphics[width=0.31\linewidth]{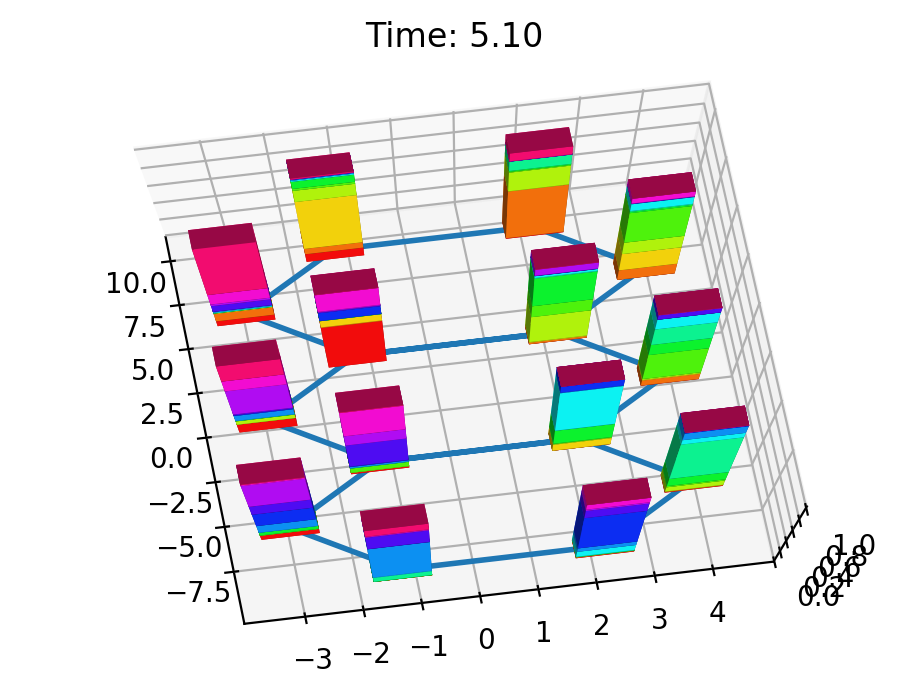} \\
				(a) & (b) \\
				\includegraphics[width=0.31\linewidth]{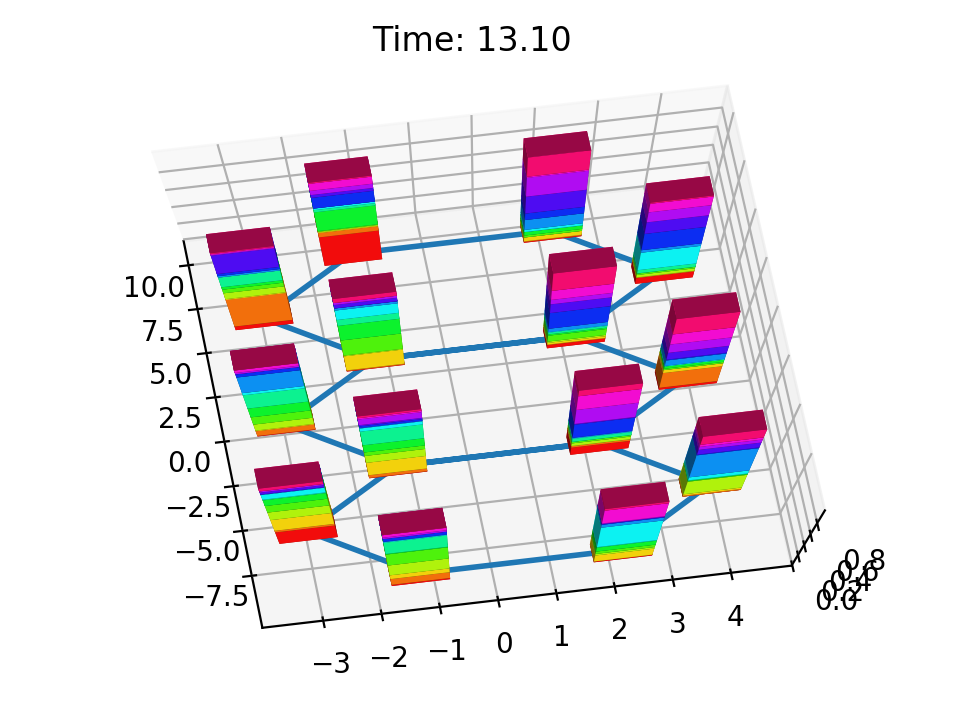} &
				\includegraphics[width=0.31\linewidth]{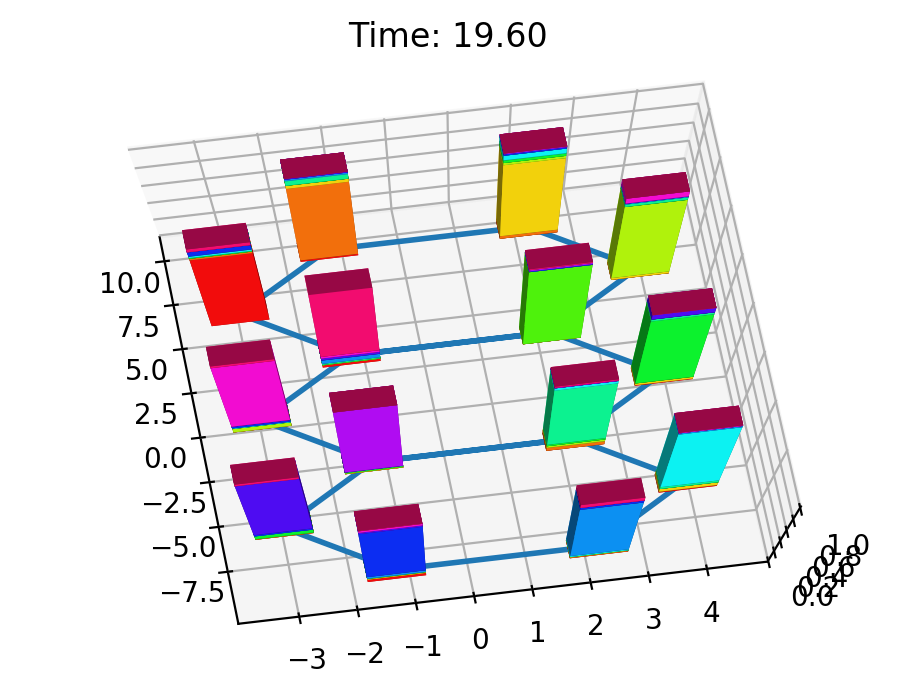} \\
				(c) & (d)\\
			\end{tabular}
			\captionof{figure}{
				The evolution of the probability distributions of the different walkers in the network representing Anthracene, with the position basis superimposed on a representation of the network representing Anthracene. Each walker is represented by a different color.
				\label{fig:fig5.3}
			}
		\end{minipage}
	\end{table}
\end{widetext}

Unlike the behavior for Benzene, the walkers on the network representing Naphthalene do not display any oscillatory behavior, and only partially localize in time. There is significantly more delocalization as compared to Benzene, however, the total probability of finding an electron at any position point is always unity. This is clearly illustrated in Fig.~\ref{fig:fig5.2}.

Similar to the case of Napthalene, the electrons in the case of Anthracene and Phenanthrene also do not have any oscillatory behavior. A few snapshots of the state of the two systems are shown in Figs.~\ref{fig:fig5.3} and \ref{fig:fig5.4}. It may be seen qualitatively by inspection that the walkers in the case of Phenanthrene (Fig.~\ref{fig:fig5.4}) display significantly more mixing in the position basis than the case of Anthracene (Fig.~\ref{fig:fig5.3}), thus alluding to the fact that Phenanthrene is more stable than Anthracene. This idea is further developed and illustrated in Fig.~\ref{fig:fig5.8}.

\begin{widetext}
	\begin{table}[H]
		\centering
		\begin{minipage}[c]{0.85\textwidth}
			\centering
			\begin{tabular}{cc}
				\includegraphics[width=0.4\linewidth]{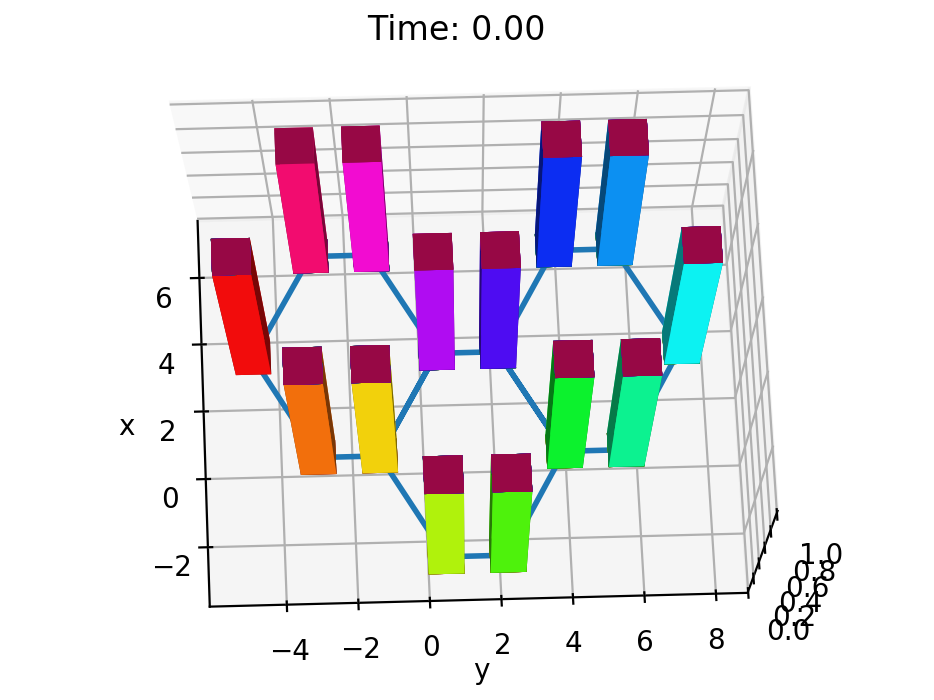} &
				\includegraphics[width=0.4\linewidth]{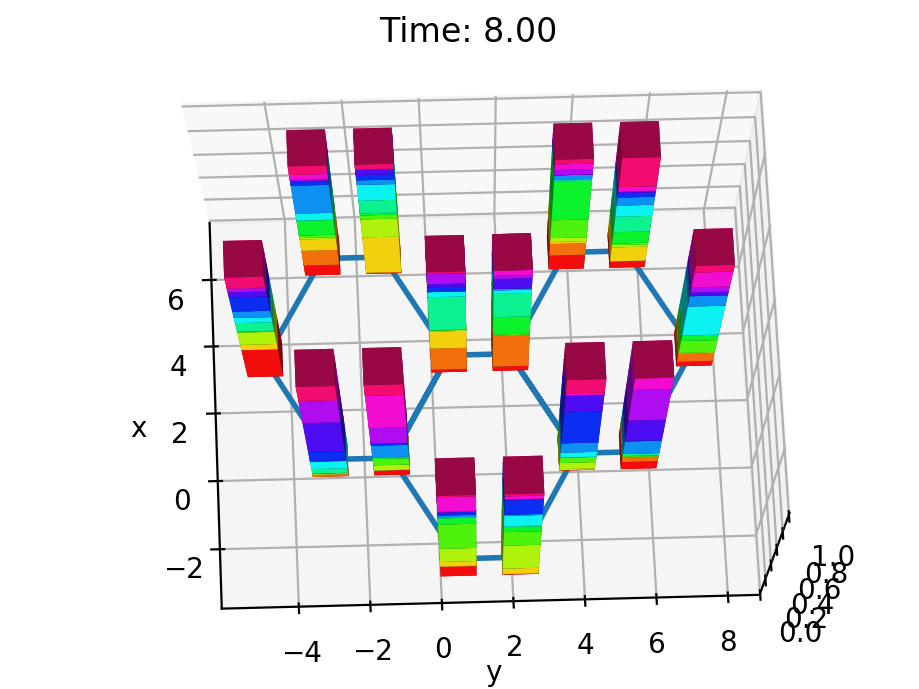} \\
				(a) & (b) \\
				\includegraphics[width=0.4\linewidth]{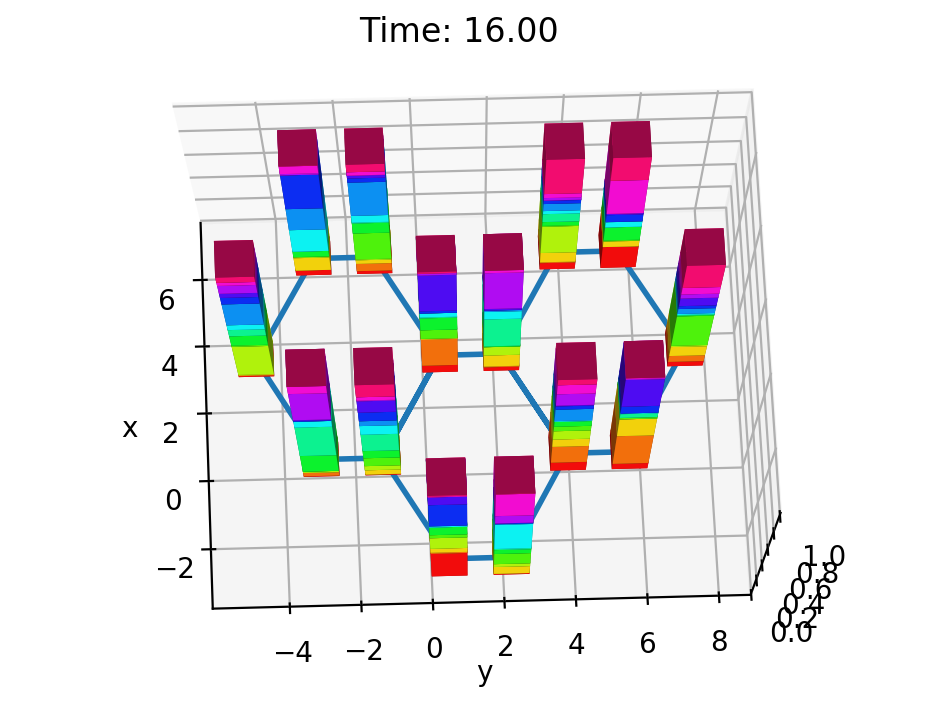} &
				\includegraphics[width=0.4\linewidth]{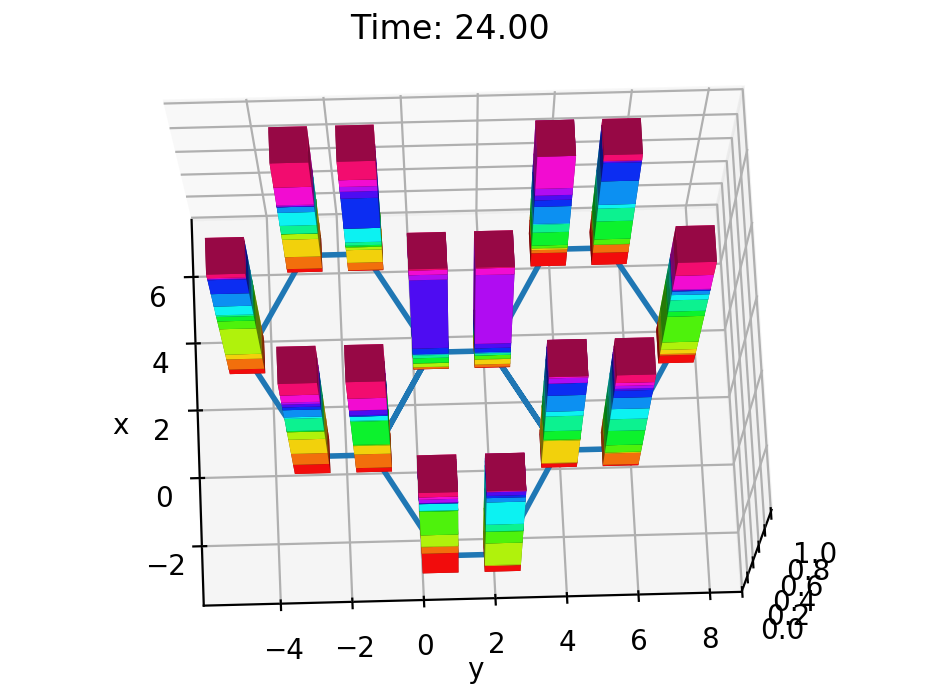} \\
				(c) & (d)\\
			\end{tabular}
			\captionof{figure}{
				The evolution of the probability distributions of the different walkers in the network representing Phenanthrene, with the position basis superimposed on a representation of the network representing the Phenanthrene molecule. Each walker is represented by a different color.
				\label{fig:fig5.4}
			}
		\end{minipage}
	\end{table}
\end{widetext}

\subsection{Reactivity of sites}

In this subsection, we look at the results obtained upon application of the DTQW-based algorithm. This analysis takes into consideration a single quantum walker diffusing over the entire network through a directed variant of the DTQW, and generating an ordered arrangement, i.e. ranking, of the nodes. This ordered arrangement is based on the amount of information passing through each node. The algorithm used is invariant of the starting position of the walker in case of a finite network over sufficient run-times, and thus it is only necessary to run it for a single walker. 

\begin{widetext}
	\begin{table}[!h]
		\centering
		\begin{minipage}[c]{0.85\textwidth}
			\centering
			\begin{tabular}{cc}
				\includegraphics[width=0.45\linewidth]{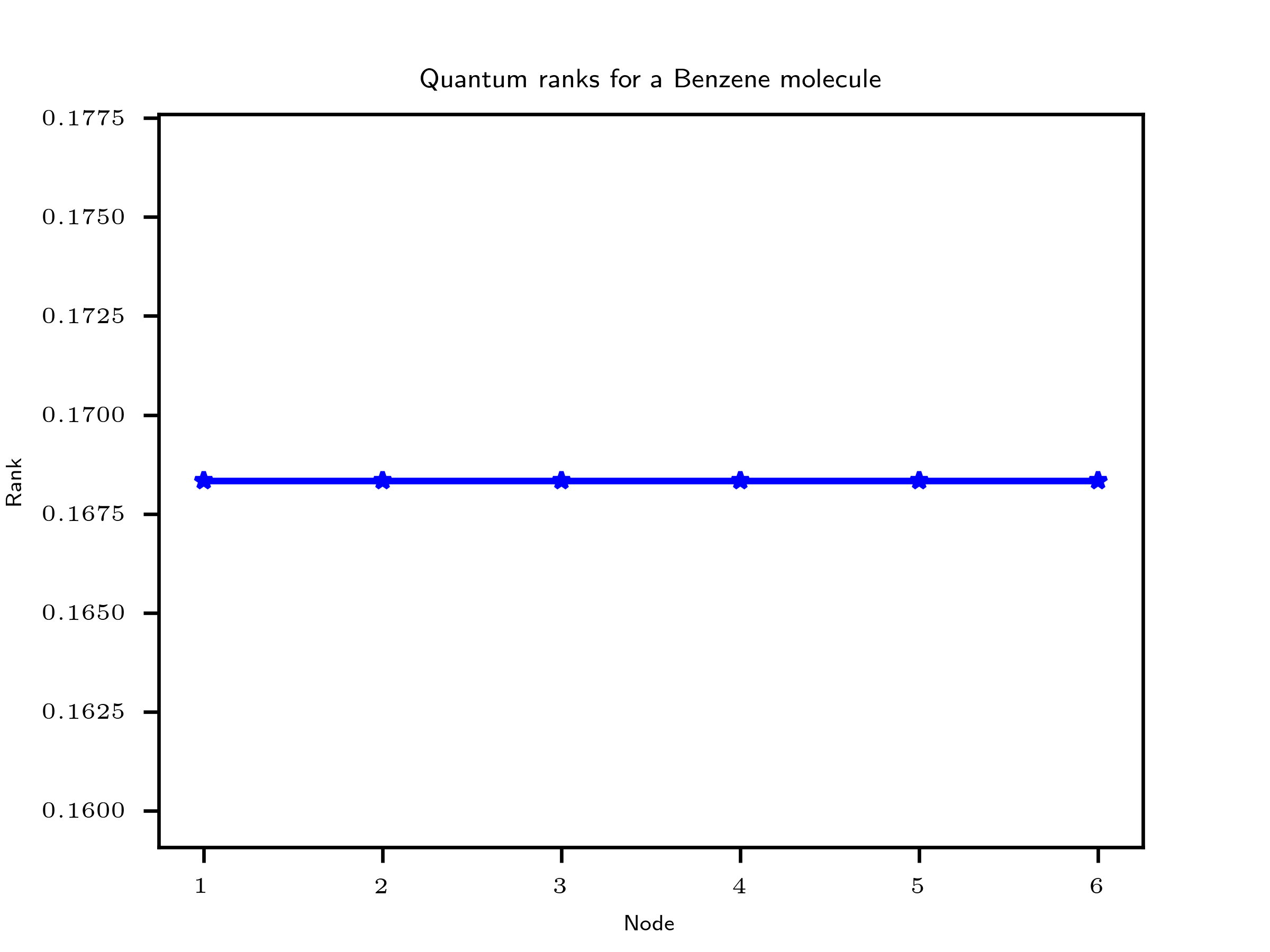} &
				\includegraphics[width=0.45\linewidth]{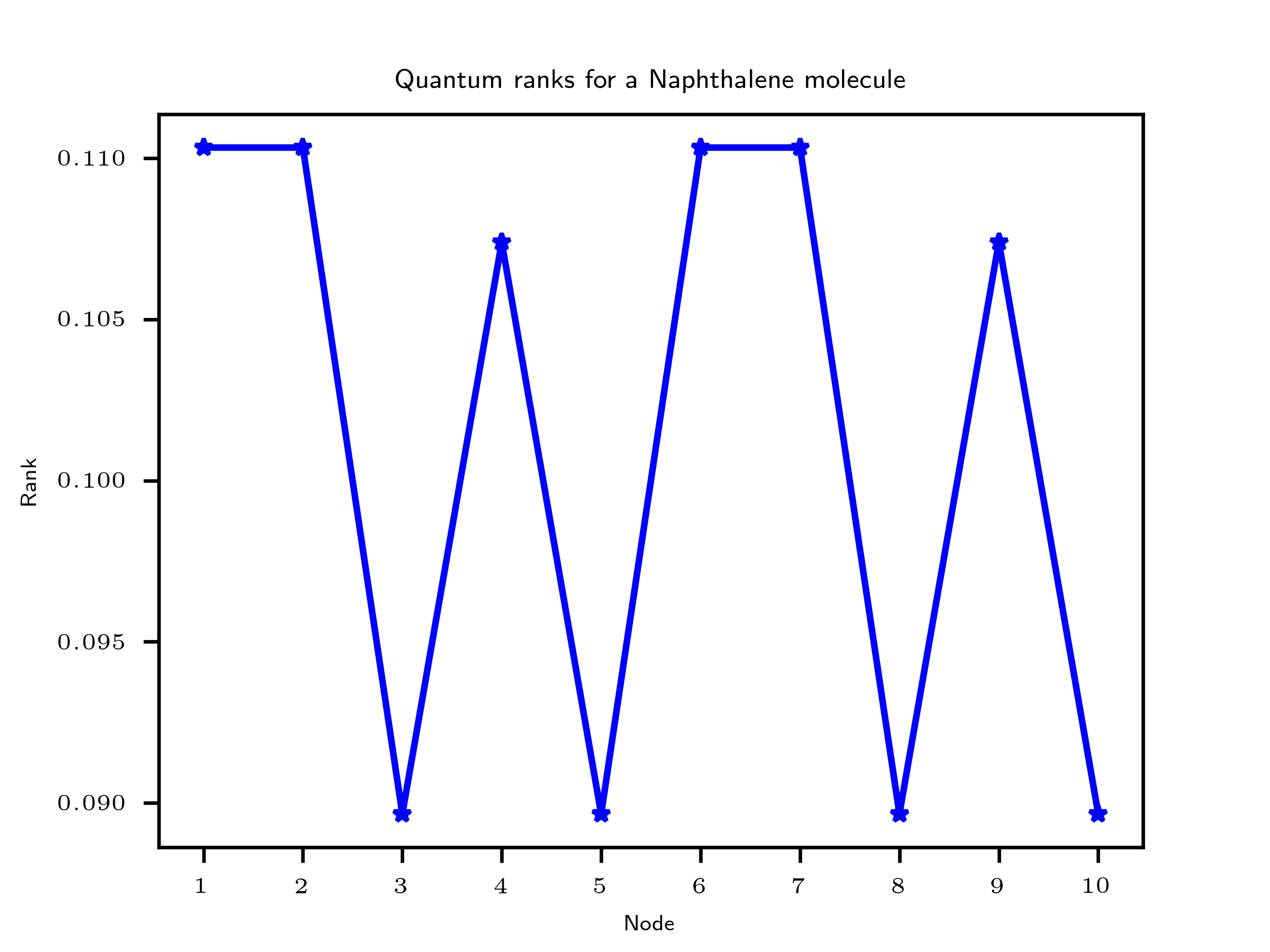} \\
				(a) & (b) \\
				\includegraphics[width=0.45\linewidth]{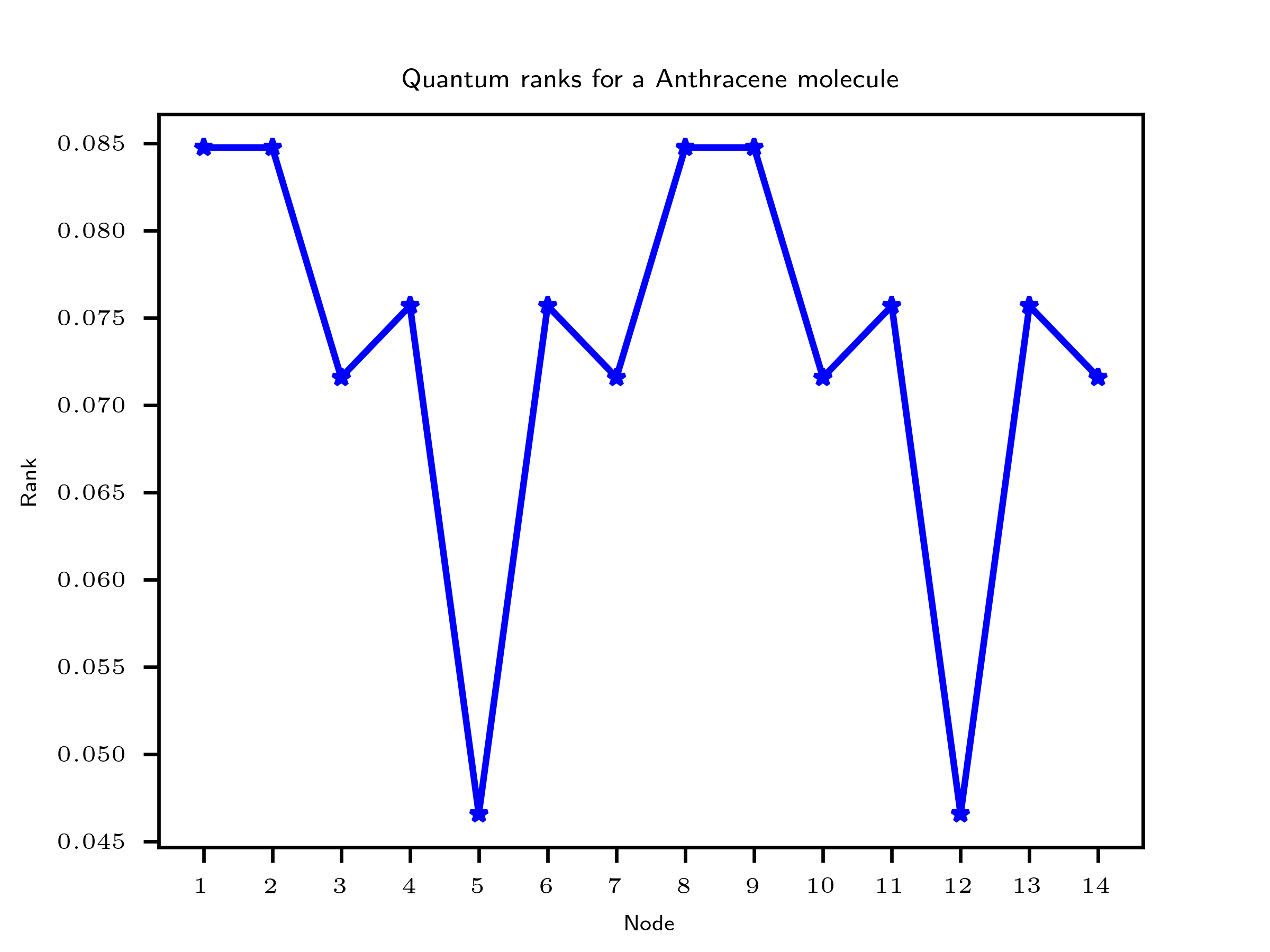} &
				\includegraphics[width=0.45\linewidth]{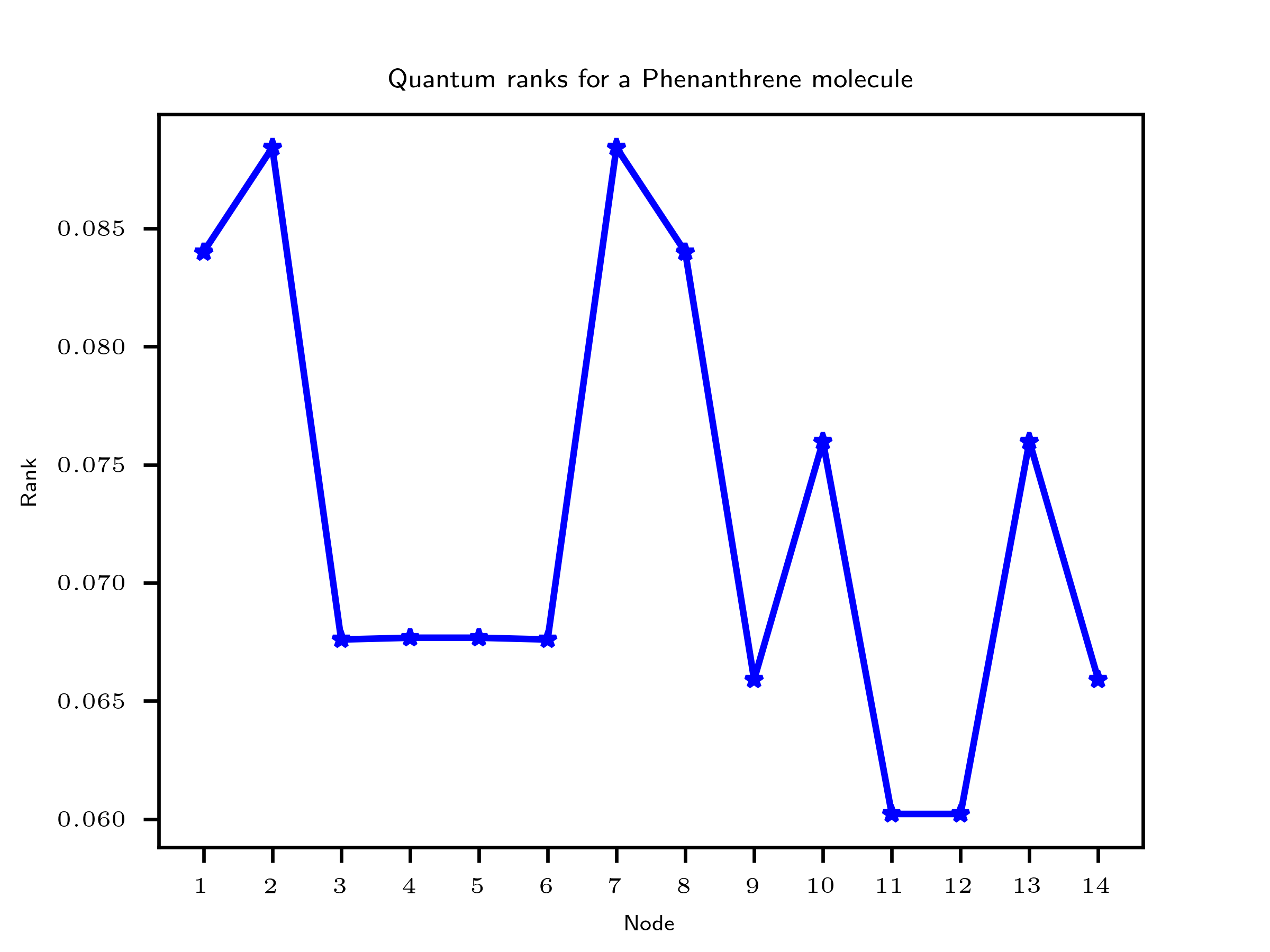} \\
				(c) & (d)\\
			\end{tabular}
			\captionof{figure}{
				Results of applying the node-ranking algorithm from \cite{CMC20} on each of the networks representing the different molecules. The sites which are equivalent have the same ranks, and therefore the same reactivity. The plots in (a), (b), (c), (d) show the results of the algorithm when applied to networks representing Benzene, Naphthalene, Anthracene and Phenanthrene molecules, respectively.
				\label{fig:fig5.5}
			}
		\end{minipage}
	\end{table}
\end{widetext}

The sites in Fig.~\ref{fig:fig5.5} are numbered according to the scheme illustrated in Fig.~\ref{fig:fig3.1}. From Fig.~\ref{fig:fig5.5}(a), it is apparent that each of the Benzene sites are equivalent, and hence an electrophilic substitution is equally likely to occur at any of the carbon atoms. This is the same result as obtained by symmetry -- the Benzene molecule has $D_{6h}$ symmetry, and thus each of its carbon atoms and the bonds in the network are equivalent. This is not the same case, however, for Naphthalene (three unique sites, $D_{2h}$ symmetry group), Anthracene (four unique sites, symmmetry group $D_{2h}$) and Phenanthrene (six unique sites, $C_{2v}$ symmetry group). In each case, a lower rank for a particular implies a higher probability of an electrophilic substitution occurring at that site. Physically, this corresponds to the fact that the position most likely to be substituted by an electrophile has the least information passing through it. The reaction, therefore, tends to occur via the pathway causing the least loss of information in the molecule, as is expected by the second law of thermodynamics.

These observations are also in line with the fact that an electrophile would preferentially create a substituted aromatic compound at site $C3$ for Naphthalene (or equivalently, at sites $C5,C8$ or $C10$). It also explains naturally the tendency of Anthracene to form substituted compounds at equivalent locations $C5$ and $C12$, as well as the high tendency of electrophilic substitutions at $C11$ and $C12$ in the case of Phenanthrene.

\subsection{Delocalization modes}

In this section, we take a look at the the variation of maximum probability of 
an electron to exist at a particular site (MAXP), for each unique site in the 
molecule. For the Benzene molecule, only the vertex $1$ is unique, and all 
others are equivalent sites. In case of the Napthalene, vertices $4,6$ and $10$ 
are unique. Anthracene and Phenanthrene, despite having the same number of 
sites, have different numbers of unique sites due to their structures. In case 
of Anthracene, sites $2,5,11$ and $14$ are unique, while in case of 
Phenanthrene, the unique sites are the ones labeled $1,2,3,9,10$ and $13$. In 
Fig.~\ref{fig:fig5.6}, we show a plot showing how the quantity MAXP varies with 
time. We have only plotted this for positions that are unique. Every equivalent 
position has the same curve, and the same mean. The values are sampled at a 
time interval of every $0.01$ units.

From Fig.~\ref{fig:fig5.6}, we observe that the plot for Benzene 
(Fig.~\ref{fig:fig5.6}(a)) shows a periodic behavior, while the others do not 
show any periodicity as such. The mean values for each position are plotted in 
Fig.~\ref{fig:fig5.7}. Through a representation of the mean value of MAXP 
measured over time, Fig.~\ref{fig:fig5.7} essentially presents a way of looking 
at the bond delocalization mode of the different bonds in the molecule. A 
vertex with a high mean value implies there is at least one higher-order bond 
at the corresponding site on the molecule, while a vertex with a lower value 
implies all the bonds at the corresponding site have delocalized. 

\begin{widetext}
	\begin{table}[H]
		\begin{minipage}[c]{\textwidth}
			\centering
			\begin{tabular}{cc}
				\includegraphics[width=0.4\linewidth]{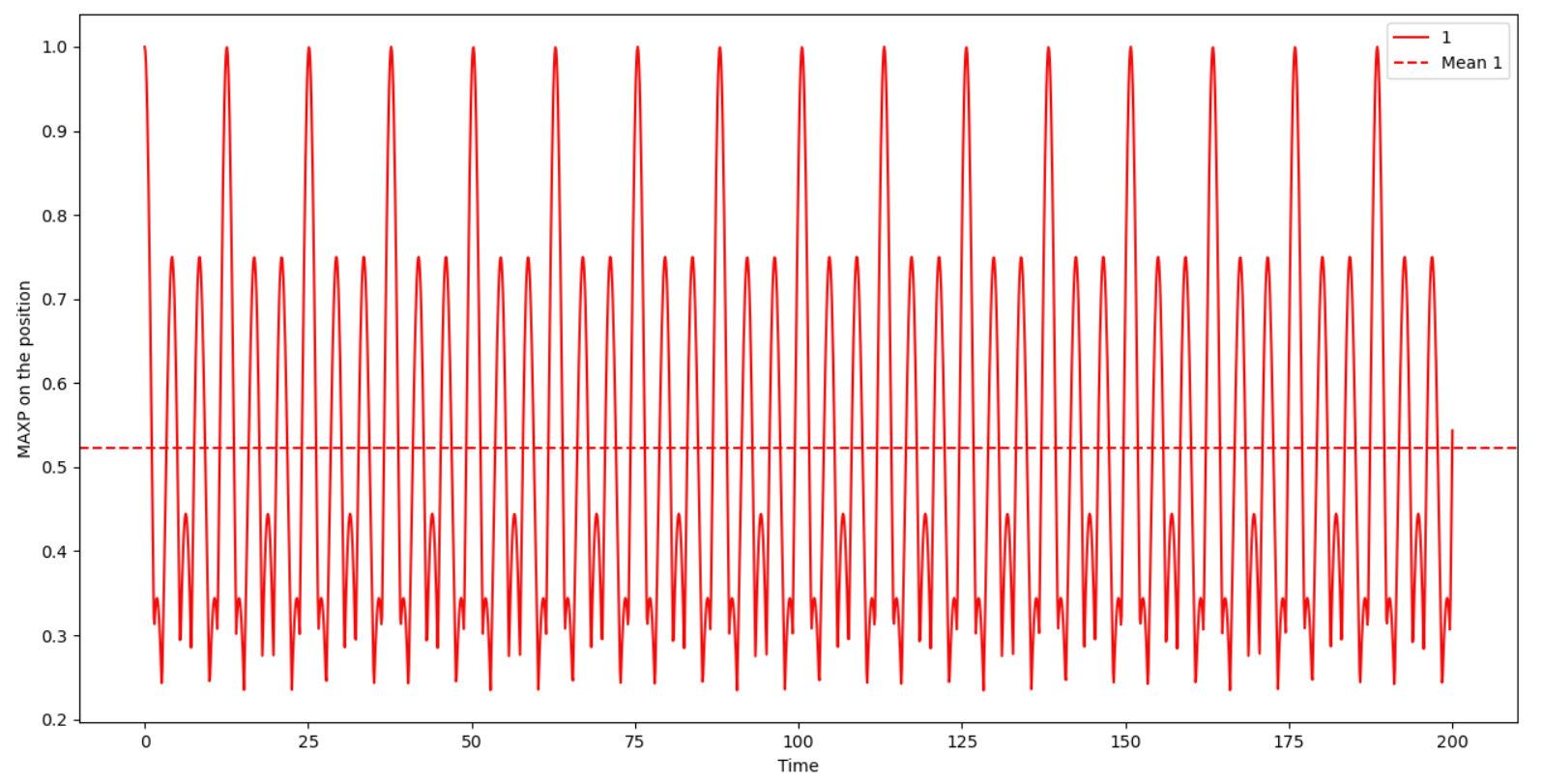} &
				\includegraphics[width=0.4\linewidth]{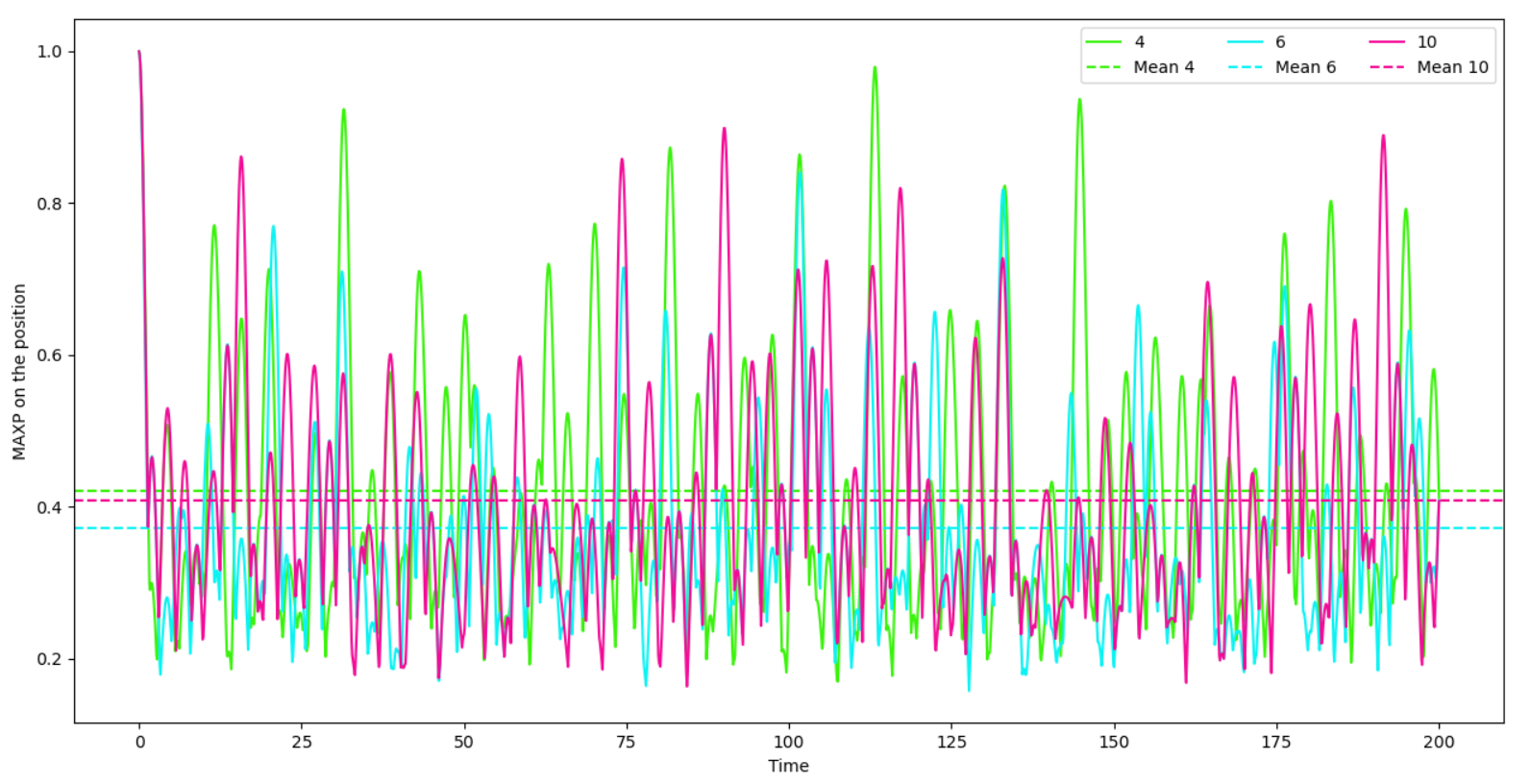} \\
				(a) & (b) \\
				\includegraphics[width=0.4\linewidth]{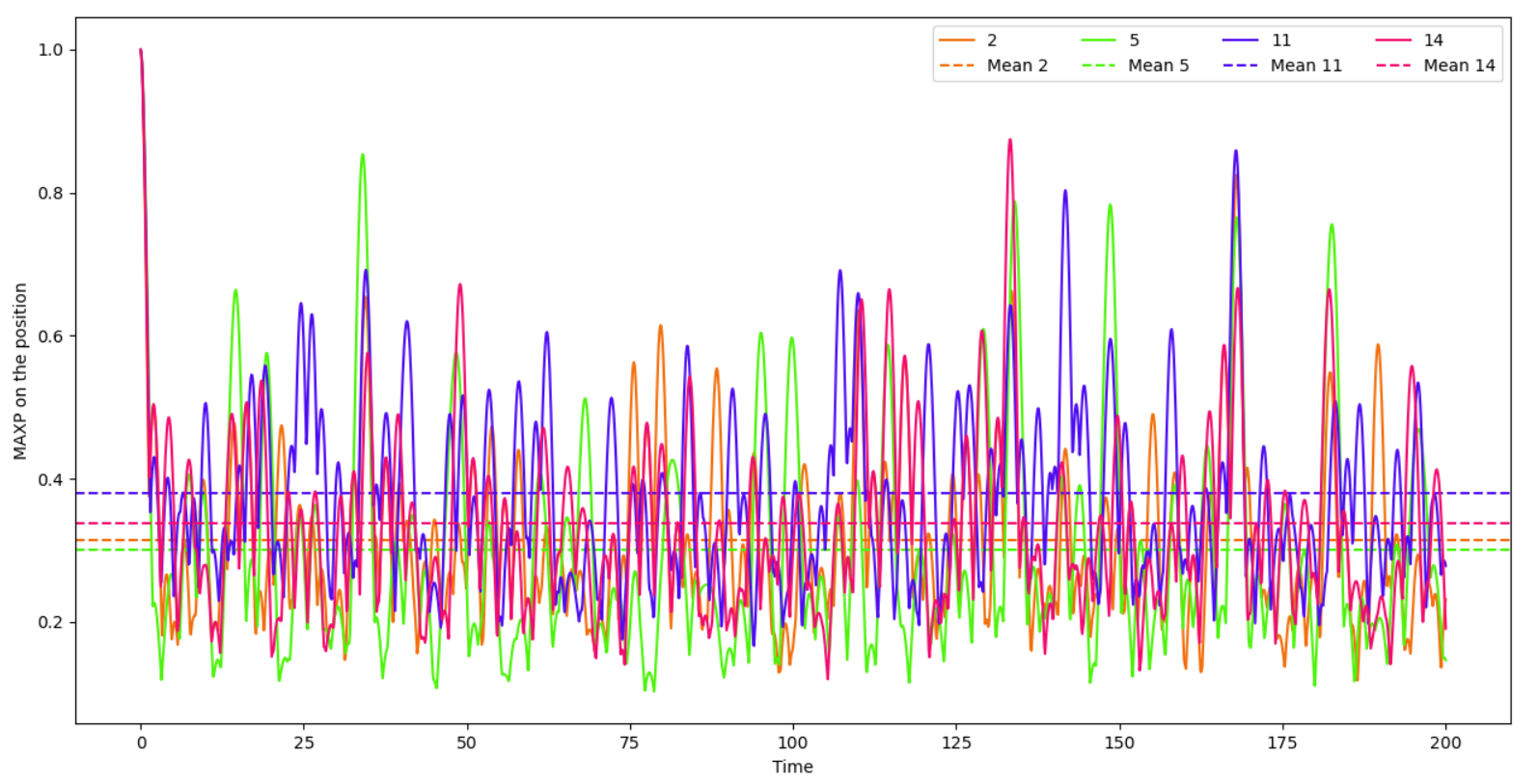} &
				\includegraphics[width=0.4\linewidth]{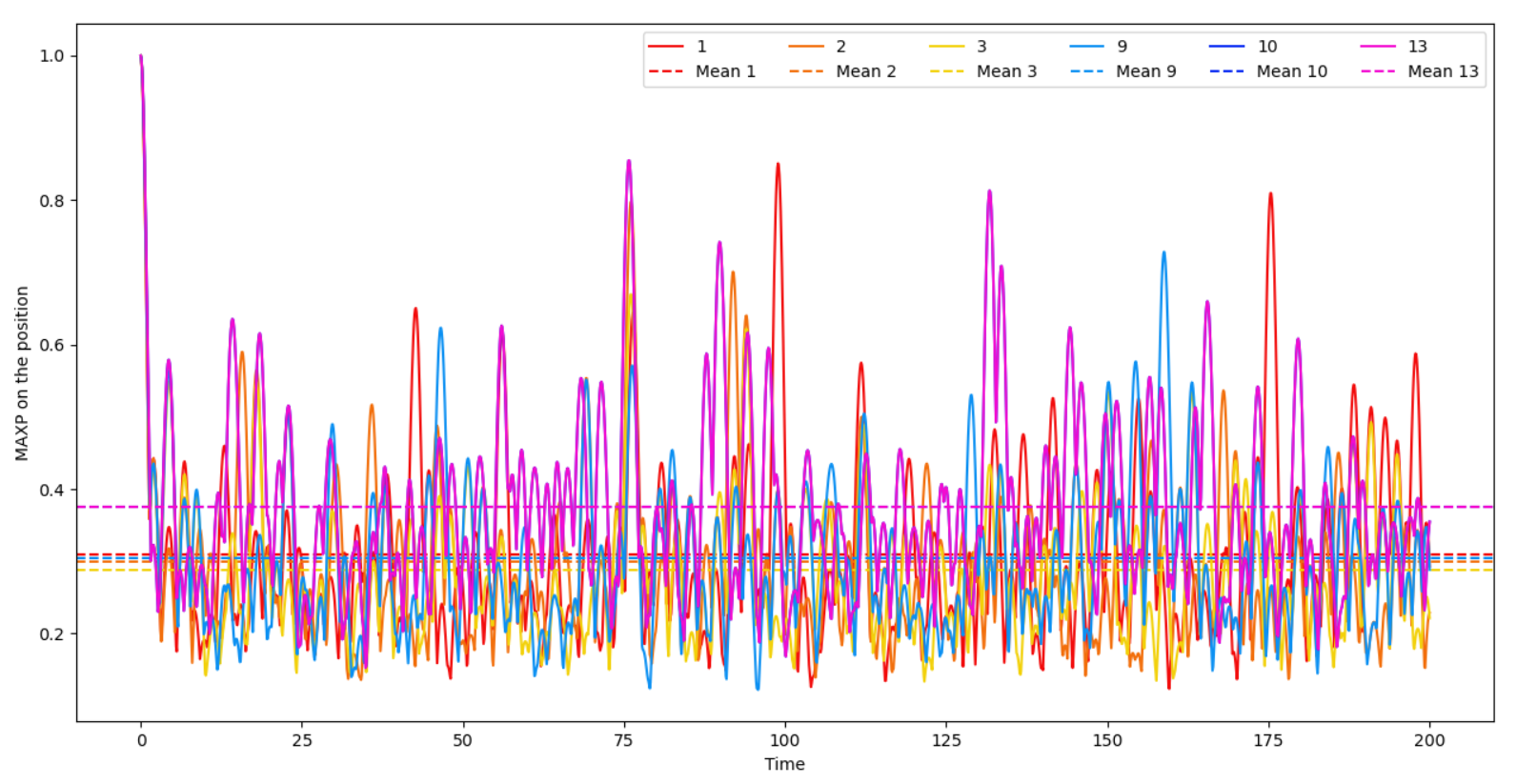} \\
				(c) & (d)\\
			\end{tabular}
			\captionof{figure}{
				Variation of MAXP with time at every node on each network considered in this work. The plots in (a), (b), (c), (d) depict the results obtained when applied to networks representing Benzene, Naphthalene, Anthracene and Phenanthrene molecules, respectively. A large amount of dynamical variation is observed, however, the mean values are plotted with dotted lines. The plot for Benzene shows a periodic variation, as expected.
				\label{fig:fig5.6}
			}
		\end{minipage}
	\end{table}
\end{widetext}

\begin{widetext}
	\begin{table}[H]
		\begin{minipage}[c]{\textwidth}
			\centering
			\begin{tabular}{cc}
				\includegraphics[width=0.4\linewidth]{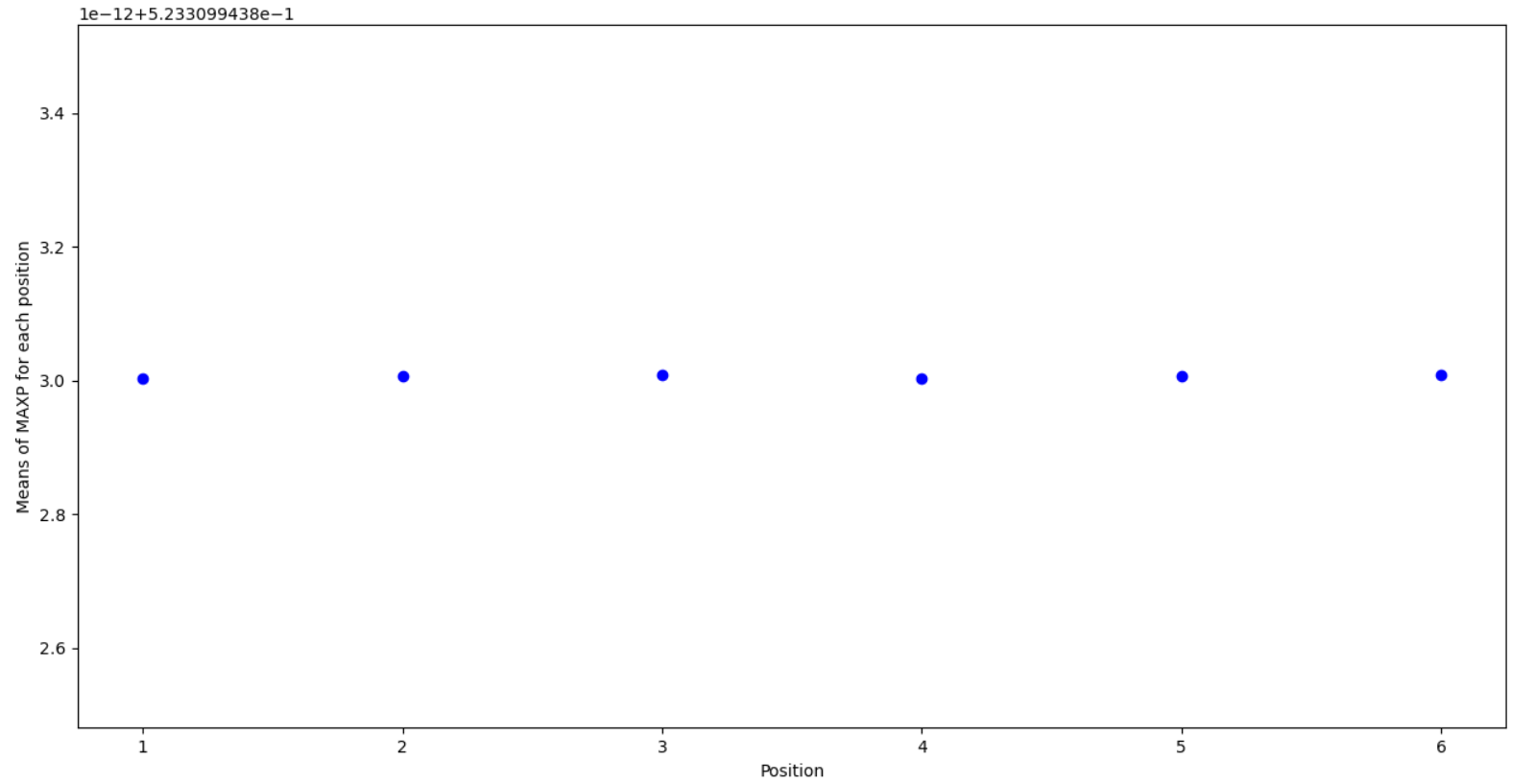} &
				\includegraphics[width=0.4\linewidth]{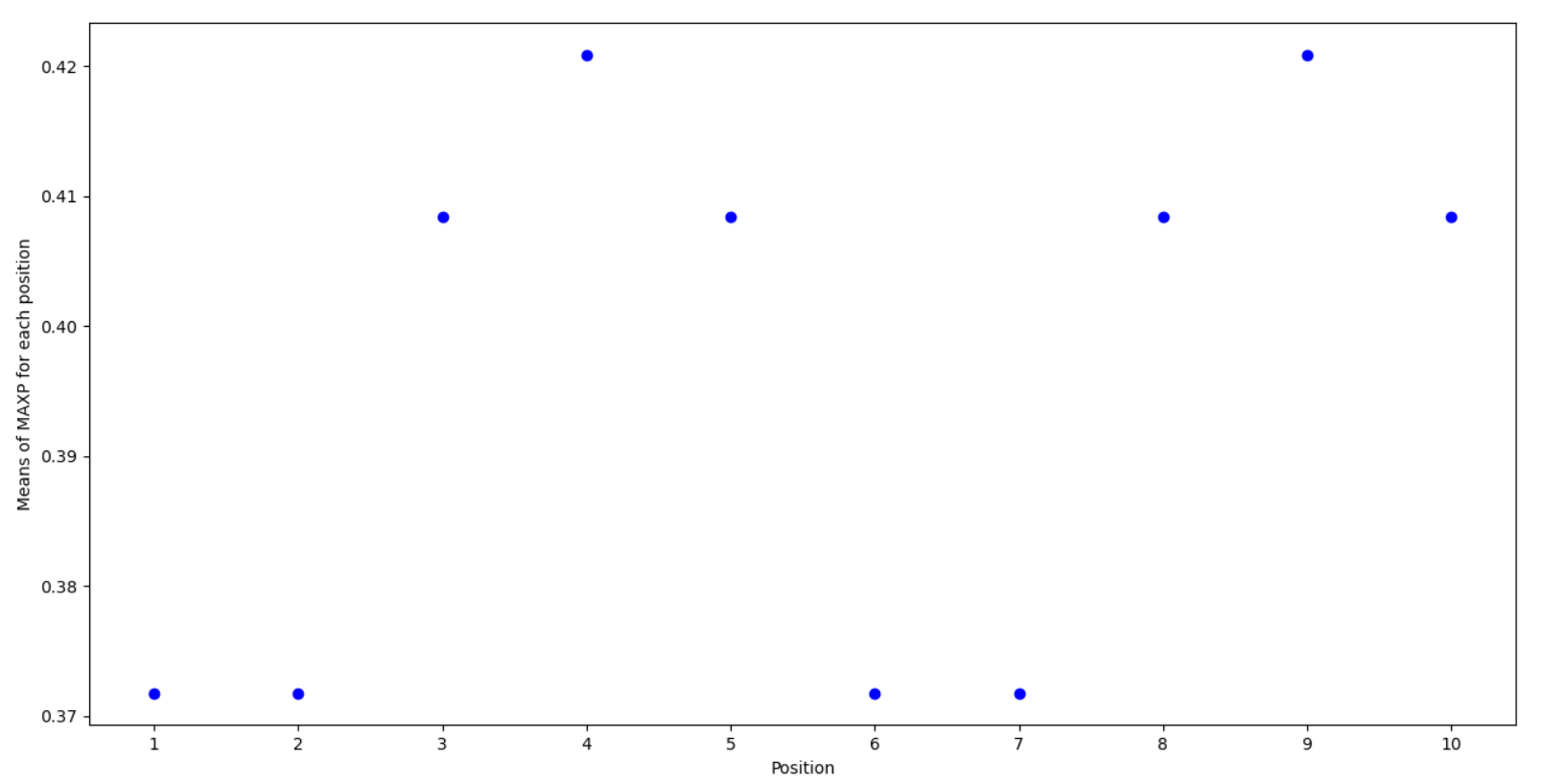} \\
				(a) & (b) \\
				\includegraphics[width=0.4\linewidth]{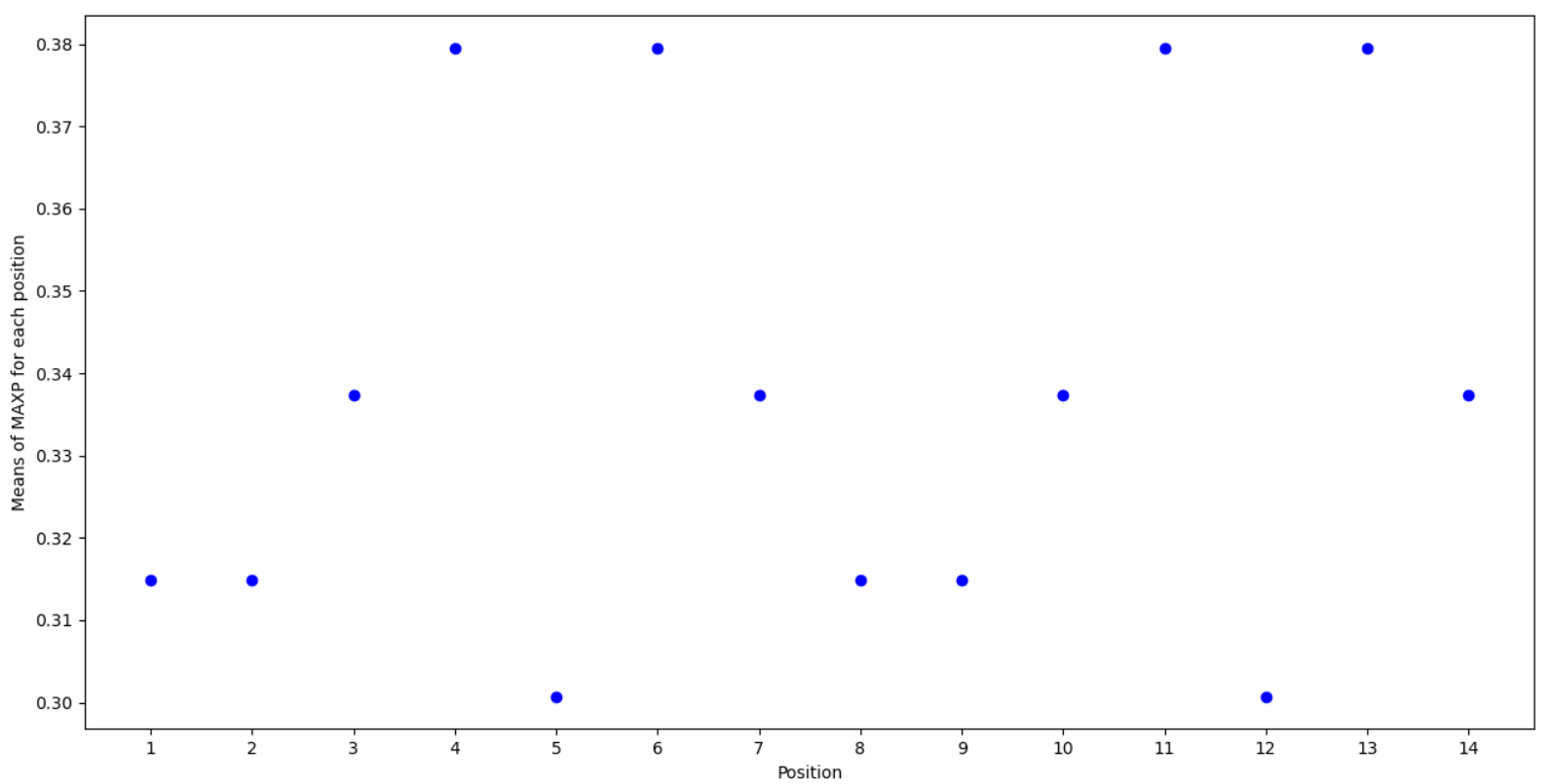} &
				\includegraphics[width=0.4\linewidth]{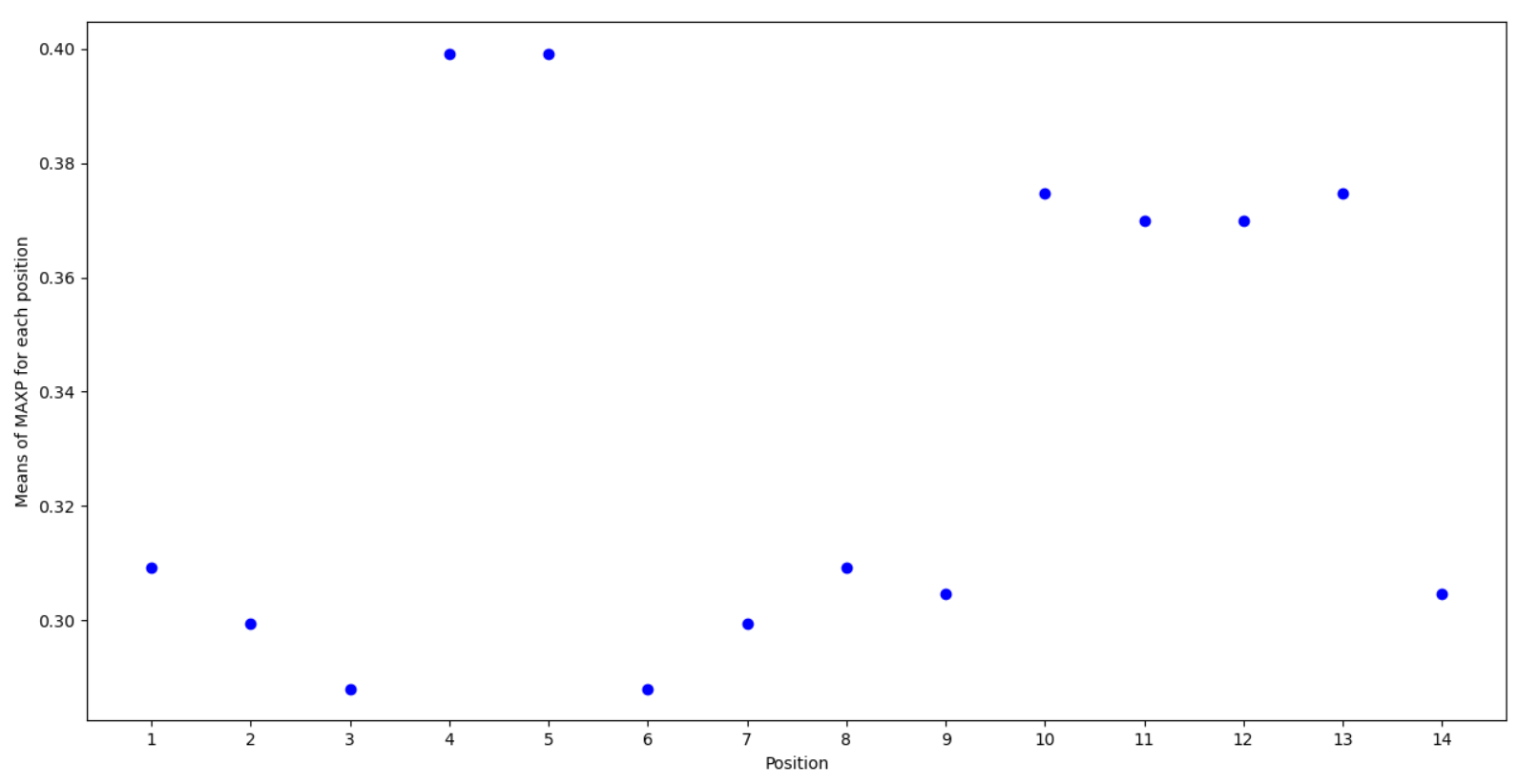} 
				\\
				(c) & (d)\\
			\end{tabular}
			\captionof{figure}{
				Mean of MAXP for every node of each network considered in this 
				work, averaged over 200 units of time, sampled at every 0.01 
				units. The plots in (a), (b), (c), (d) show the results 
				obtained when applied to networks representing Benzene, 
				Naphthalene, Anthracene and Phenanthrene molecules, 
				respectively. As expected, the plot for Benzene shows that 
				every site is equivalent.
				\label{fig:fig5.7}
			}
		\end{minipage}
	\end{table}
\end{widetext} %

This enables us to verify the delocalization mode for each of the molecules. Benzene has only one possible pattern shown in Fig.~\ref{fig:fig4.1}(a), but Naphthalene prefers to exist in the pattern depicted in Fig.~\ref{fig:fig4.1}(c). Similarly, Anthracene prefers to exist in a mixture of the patterns shown in Fig.~\ref{fig:fig4.1}(d) and Fig.~\ref{fig:fig4.1}(f), however, the pattern in Fig.~\ref{fig:fig4.1}(d) is slightly more dominant. In the case of Phenanthrene, the dominant arrangement is that of a biphenyl unit connected by a bridge, illustrated in Fig.~\ref{fig:fig4.1}(i), mixed with the slightly less preferred peripherally delocalized pattern shown in  Fig.~\ref{fig:fig4.1}(g). 

\begin{widetext}
	\begin{table}[H]
		\begin{minipage}[c]{\textwidth}
			\centering
			\begin{tabular}{cc}
				\includegraphics[width=0.45\linewidth]{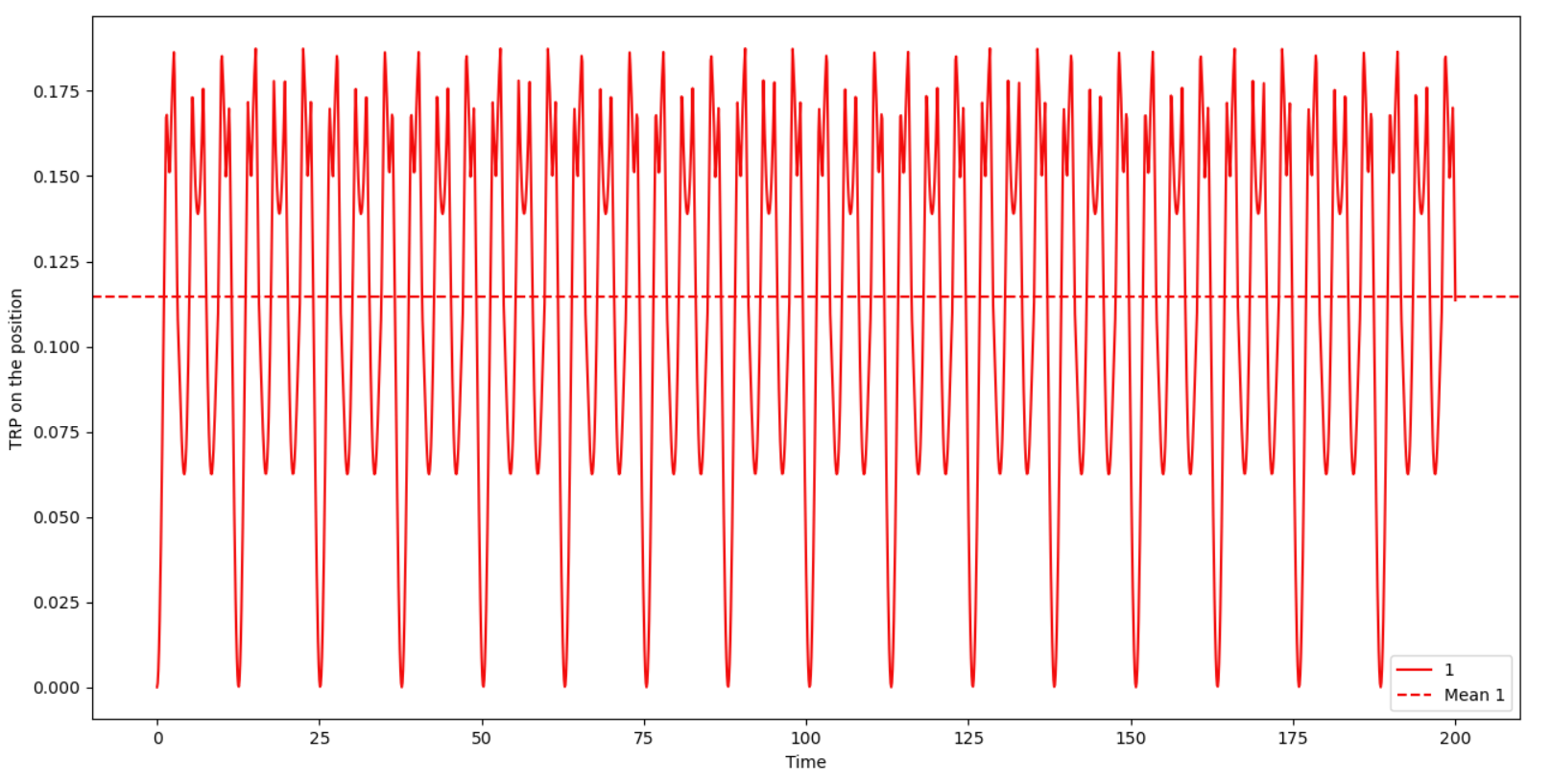} &
				\includegraphics[width=0.45\linewidth]{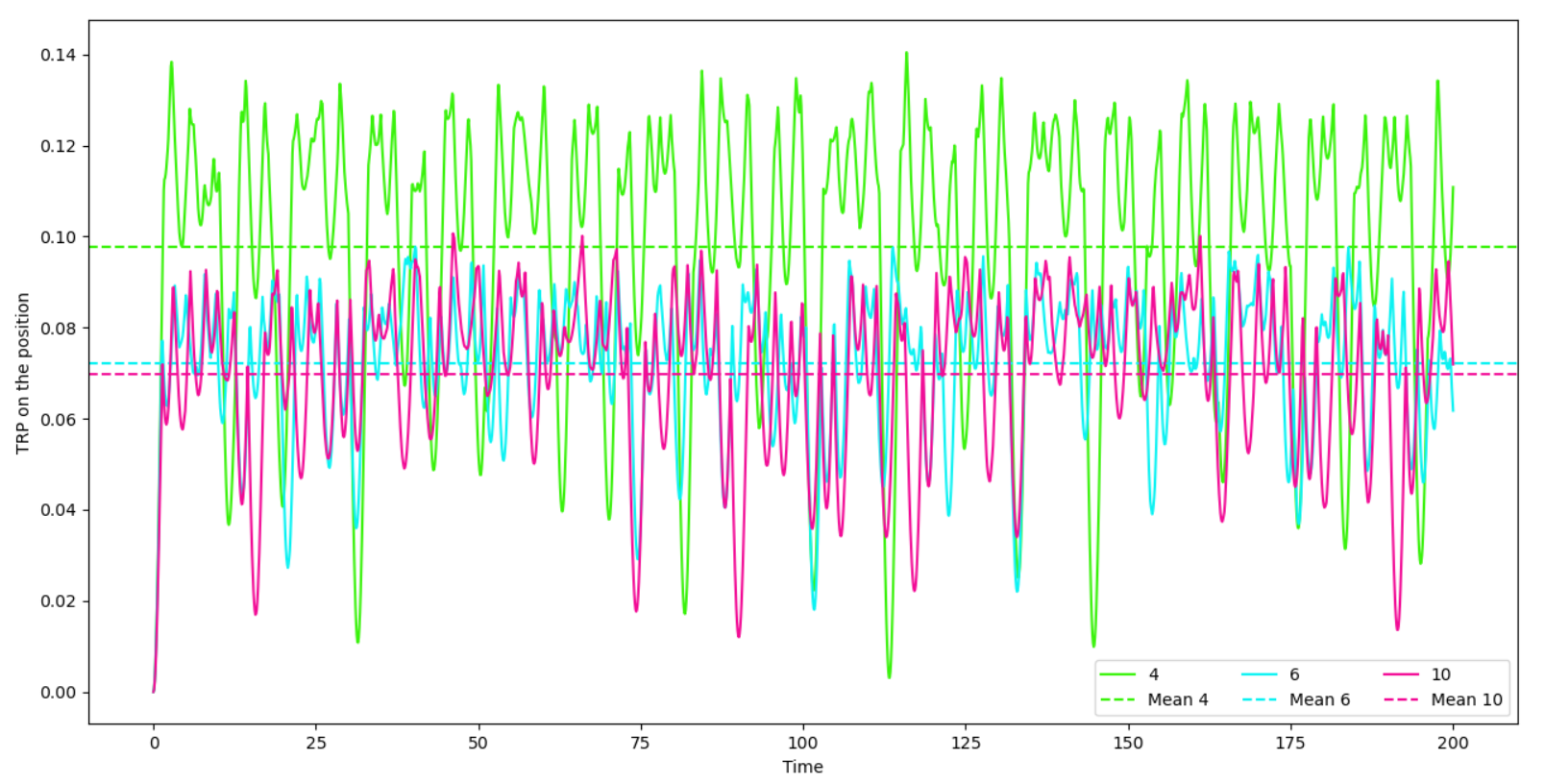}
				 \\
				(a) & (b) \\
				\includegraphics[width=0.45\linewidth]{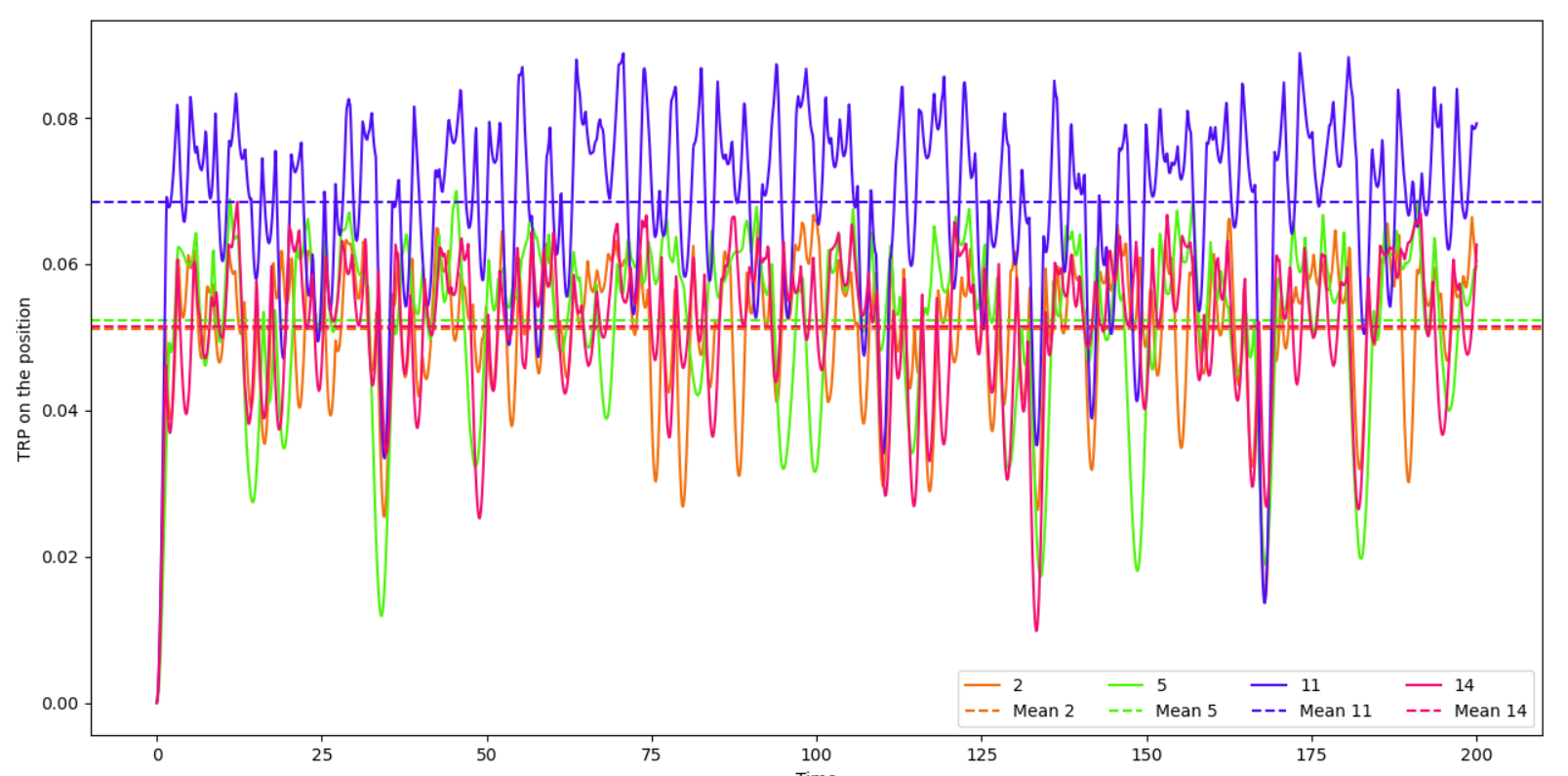} 
				&
				\includegraphics[width=0.45\linewidth]{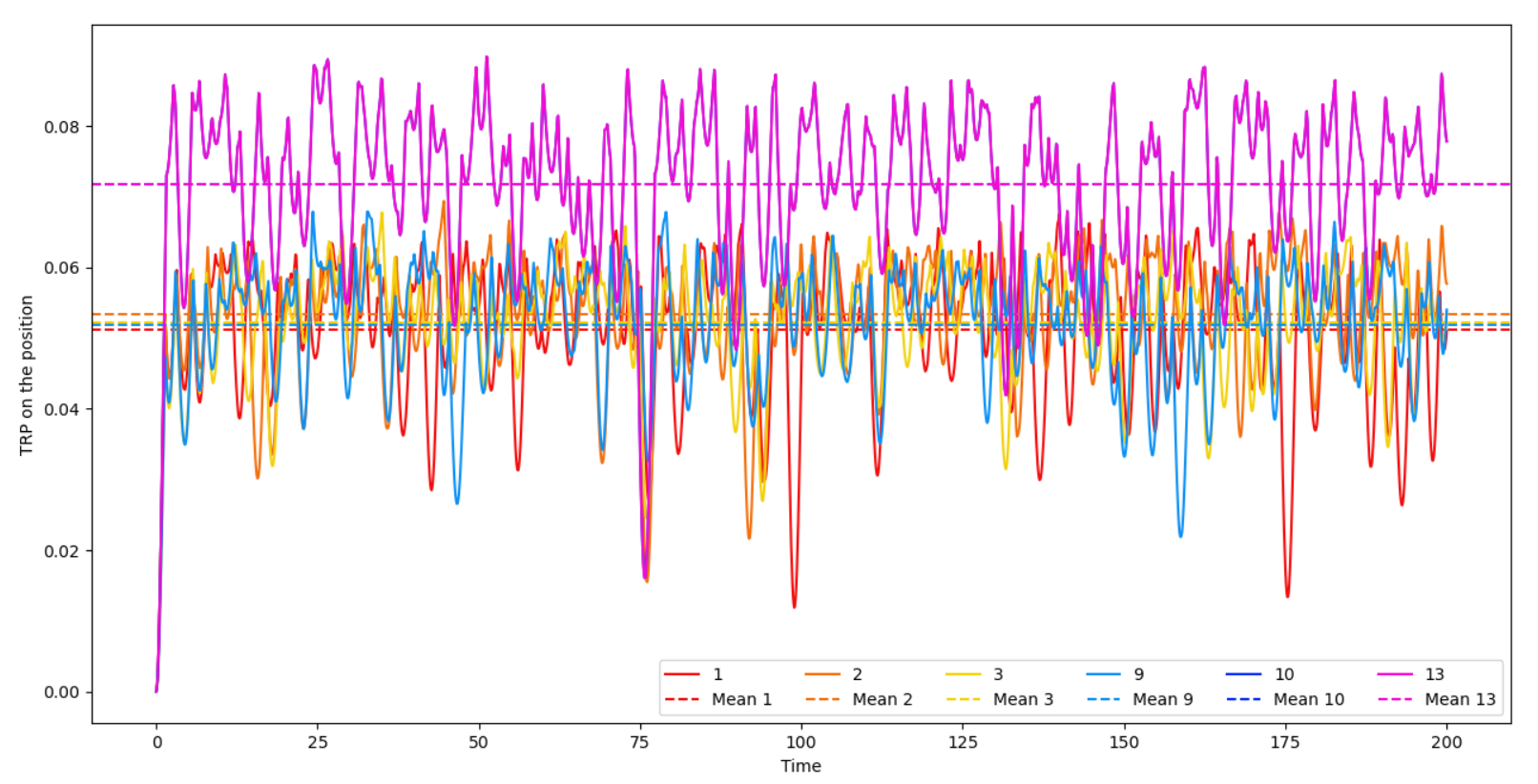}
				 \\
				(c) & (d)\\
			\end{tabular}
			\captionof{figure}{
				TRP of the probability of an electron to exist at a particular 
				site on each molecule, plotted for 200 units of time, sampled 
				at every 0.01 units. The plots in (a), (b), (c), (d) depict the 
				results obtained when applied to networks representing Benzene, 
				Naphthalene, Anthracene and Phenanthrene molecules, 
				respectively. The dotted line shows the mean value of the plot, 
				averaged over time.
				\label{fig:fig5.8}
			}
		\end{minipage}
	\end{table}
\end{widetext}

\subsection{Order of stability}

The final metric that we have considered is the truncated mean of the probabilities of the electrons to exist at various positions, also known as TRP. The plots of TRP for unique (i.e. inequivalent) positions for each molecule are shown in Fig.~\ref{fig:fig5.8}. Just as the case of MAXP, a lot of dynamical variation can be observed, however, a higher mean value here implies that electrons do not fully localize at that position, even for a very small time, and the position is a part of the delocalized cloud. A higher TRP over all the positions also implies the species is more stable overall as all bonds tend to have a higher degree of delocalization, and a consequently a higher resonance energy. It thus naturally gives rise to the known stability order of the four aromatic molecules considered, i.e. $\text{Benzene} > \text{Naphthalene} \sim \text{Phenanthrene} > \text{Anthracene}$.\\ \\

\clearpage
\section{Conclusion}
\label{sec:conc}

We have studied the structure and properties of four benzoid polycyclic aromatic hydrocarbons using quantum walks. We have characterized the evolution of the probability of finding electrons at different points, and developed metrics with the help of which we are able to qualitatively understand the structure of these molecules. We also use some of the developed metrics to characterize the chemical properties of these molecules, namely, the characterization of the electron-rich sites in the structure, which are preferentially targeted by electrophiles in solution. We are also able to perform a relative stability analysis of these species, and our results agree with the previously established results from previous chemical studies.

This work also proposes a formalism in the analysis of aromatic compounds, wherein the quantum walk is the fundamental physical process by which electrons diffuse in the delocalized $\pi$-electron cloud. This has applications in simulations of chemical reactions via quantum simulators and/or quantum computers capable of realizing quantum walks on networks. For smaller molecules, the calculations may be done on a classical computer as well, making this formalism accessible to both classical and quantum computing paradigms. We aim to extend this analysis to non-planar molecules, as well as substituted aromatic molecules. {This approach also prompts the use of machine learning techniques, such as deep neural networks, in order to calculate the bond orders by heuristic data such as the one generated by the node-ranking algorithm. }


\section*{Acknowledgement}
\label{sec:ack}

We acknowledge support from the Interdisciplinary Cyber Physical Systems (ICPS) programme of the Department of Science and Technology, Government of India. Grant No: DST/ICPS/QuST/Theme-1/2019/1.



\end{document}